\documentclass[english,onecolumn]{IEEEtran}
\usepackage[T1]{fontenc}
\usepackage[latin9]{inputenc}
\usepackage{float}
\usepackage{units}
\usepackage{amsthm}
\usepackage{amssymb}
\usepackage{graphicx}
\usepackage{setspace}
\makeatletter
\theoremstyle{plain}
\newtheorem{thm}{\protect\theoremname}
\theoremstyle{definition}
\newtheorem{defn}[thm]{\protect\definitionname}
\theoremstyle{remark}
\newtheorem{rem}[thm]{\protect\remarkname}
\theoremstyle{plain}
\newtheorem{lem}[thm]{\protect\lemmaname}

\makeatother

\usepackage{babel}
\providecommand{\definitionname}{Definition}
\providecommand{\lemmaname}{Lemma}
\providecommand{\remarkname}{Remark}
\providecommand{\theoremname}{Theorem}

\begin{document}

\title{The Wideband Slope of Interference Channels: The Small Bandwidth
Case}

\author{Minqi Shen, Anders H{\normalsize {\o}st-Madsen}%
\thanks{The authors are with the Department of Electrical Engineering, University
of Hawaii Manoa, Honolulu, HI 96822 (e-mail: \{minqi,ahm\}@hawaii.edu.
This work was supported in part by NSF grant CCF 1017823. This paper
was presented in part at the 49th annual Allerton Conference on Communication,
Control and Computing, September 2011 (Urbana-Champaign, IL).%
}}

\maketitle
\global\long\def\snr{\mathrm{SNR}}
\global\long\def\kuser{K-\mathrm{user}}
\global\long\def\trace{\mathrm{Tr}}
\global\long\def\var{\mathrm{var}}
\global\long\def\cov{\mathrm{cov}}
\global\long\def\ebno{\frac{E_{b}}{N_{0}}}
\global\long\def\slope{\mathcal{S}}
\global\long\def\mod{\mathrm{mod}}
\global\long\def\sinc{\mathrm{sinc}}

\global\long\def\ebnomin{\left.\frac{E_{b}}{N_{0}}\right|_{\min}}
\global\long\def\ebnominsum{\left.\frac{E_{b}}{N_{0}}\right|_{\min_{0}}}

\begin{abstract}
This paper studies the low-SNR regime performance of a scalar complex
$K$-user interference channel with Gaussian noise. The finite bandwidth
case is considered, where the low-SNR regime is approached by letting
the input power go to zero while bandwidth is small and fixed. We
show that for all $\delta>0$ there exists a set with non-zero measure
(probability) in which the wideband slope per user satisfies $\slope_{0}<\nicefrac{2}{K}+\delta$.
This is quite contrary to the large bandwidth case \cite{ShenAHM11IT},
where a slope of $1$ per user is achievable with probability 1. We
also develop an interference alignment scheme for the finite bandwidth
case that shows some gain. 
\end{abstract}

\section{Introduction}

This paper and the companion paper \cite{ShenAHM11IT} study the bandwidth-power
trade-off of a $K$-user interference channel in the low-$\snr$ (signal-to-noise)
regime, where explicitly
\begin{eqnarray}
\snr & \triangleq & \frac{P}{BN_{0}}.\label{eq:def snr}
\end{eqnarray}
Bandwidth and input power, two important design parameters, are related
by the function $R\left(\ebno\right)$, where $\ebno$ is the transmitted
energy per bit, and $R$ is the spectral efficiency. The concept of
the low-$\snr$ regime was introduced by S. Verdú in the 2002 paper
\cite{Ver02IT}. A system working in this regime is characterized
by very small spectral efficiency, so that the $R\left(\ebno\right)$
curve can be closely approximated by its first-order approximation,
which is determined by two measures: the minimum energy per bit $\ebnomin$
and the wideband slope $\slope_{0}$. $\ebnomin$ is the minimum transmitted
energy per bit required by reliable communication, which is generally
achieved at zero spectral efficiency; and $\slope_{0}$ is the first-order
slope of $R\left(\ebno\right)$ as $\ebno$ approaches $\ebnomin$.
These two measures are defined by 
\begin{eqnarray}
\ebnomin & = & \lim_{\snr\downarrow0}\frac{\snr}{R\left(\snr\right)}\label{eq:ebnomin def}
\end{eqnarray}
\begin{eqnarray}
\mathcal{S}_{0} & \triangleq & \lim_{\ebno\downarrow\ebno_{\min}}\frac{R\left(\ebno\right)}{10\log_{10}\ebno-10\log_{10}\ebnomin}10\log_{10}2,\label{eq:slope def}
\end{eqnarray}
Further manipulations in \cite{Ver02IT} show that $\ebnomin$ and
$\slope_{0}$ can be determined by the first and second order derivative
of $R\left(\snr\right)$ at zero $\snr$:
\begin{eqnarray}
\ebnomin & = & \frac{\log_{e}2}{\dot{R}\left(0\right)},\label{eq:ebnomin calc}
\end{eqnarray}
\begin{eqnarray}
\slope_{0} & = & -\frac{2\left(\dot{R}\left(0\right)\right)^{2}}{\ddot{R}\left(0\right)},\label{eq:slope calc}
\end{eqnarray}
where $\dot{R}\left(0\right)$ and $\ddot{R}\left(0\right)$ are the
first-order and the second-order Taylor expansion coefficients for
$\snr\rightarrow0$. $\dot{R}\left(0\right)=\left.\frac{dR\left(\snr\right)}{d\snr}\right|_{\snr=0}$
and $\ddot{R}\left(0\right)=\left.\frac{d^{2}R\left(\snr\right)}{d\snr^{2}}\right|_{\snr=0}$
if $R\left(\snr\right)$ is differentiable. 

What is interesting is that there are two distinct ways to approach
the low-$\snr$ regime, which have very different impacts on the performance
of the interference channel Although approaching the low-$\snr$ regime
by letting $B\rightarrow\infty$ is emphasized in previous papers
(hence the term {}``wideband slope''), it is not the only way. As
can be noted from the definition of SNR (\ref{eq:def snr}), SNR approaches
zero if either $B\to\infty$ or $P\to0$. Consider a point-to-point
AWGN channel with spectral efficiency
\begin{eqnarray*}
R & = & \log\left(1+\frac{P}{BN_{0}}\right).
\end{eqnarray*}
The low-$\snr$ results are based on a Taylor series of $\log(1+x)$
, as also seen by (\ref{eq:ebnomin calc}-\ref{eq:slope calc}); therefore
as long as $\snr=\frac{P}{BN_{0}}\to0$ in any manner, low-SNR results
such as minimum energy per bit and wideband slope are unchanged. The
key is that the spectral efficiency $R\to0$, not that $B\to\infty$.
For the interference channel, on the other hand, different results
are obtained depending on how the low-$\snr$ regime is approached.

In the first approach, let $B\to\infty$ while $P$ is fixed and finite.
We call this the \emph{large bandwidth regime}. In \cite{ShenAHM11IT}
we proved that in this case a wideband slope of $K$ was achievable
with probability one by using channel delays.

In the second approach, let $P\rightarrow0$ while $B$ is fixed and
finite. In this case, the rate $BR$ in bits/s must necessarily approach
0 as well, and we therefore call this the \emph{low-rate regime}.
This is the case considered in this paper, and as will be seen the
results are quite different than the the case in \cite{ShenAHM11IT}.

To put the results of this paper in context, consider the completely
symmetric channel: the channel between receiver pairs $(i,j)$ is
the same for all $1\leq i,j\leq K$, both $i=j$ and $i\neq j$. We
call this channel the $\underline{1}$-channel. The capacity of this
channel is fully known: because of the symmetry all receivers must
be able to decode all messages, and the capacity is therefore given
by the MAC (multiple access channel) bound into one of the nodes.
For this channel, FDMA (frequency division multiple access) or TDMA
(time division multiple access) is optimum, and the degrees of freedom
\cite{CadambeJafar07} is 1 ($1/K$ per user) while the wideband slope
is $2$ ($2/K$ per user). A key question is if this channel is typical.
For degrees of freedom the answer is no: the results in \cite{MotGhaKha09CoRR}
and \cite{WuShamaiVerdu12} show that the degrees of freedom is $K/2$
($\frac{1}{2}$ per user) almost everywhere for a scalar channel.
Thus, the degrees of freedom is discontinuous in $\underline{1}$,
and in fact almost everywhere. Similarly, \cite{CadambeJafar07} shows
that for time-varying channels, the degrees of freedom is $K/2$ with
probability one. In \cite{ShenAHM11IT} we proved analogously that
in the \emph{large bandwidth regime} the wideband slope is $K$(1
per user) with probability one for a line-of-sight channel. Thus,
also the wideband slope is discontinuous in $\underline{1}$ and again
in fact discontinuous with probability one.

The main result of this paper is that in the \emph{low-rate regime}
the wideband slope is upper semi-continuous in $\underline{1}$. That
is, for any $\delta>0$ there exists an open set $\tilde{\mathcal{C}}_{\delta}$
of channels so that $\underline{1}\in\mbox{cl}(\tilde{\mathcal{C}}_{\delta})$
(cl means closure) and $\mathcal{S}_{0}\leq2+\delta$ in $\tilde{\mathcal{C}}_{\delta}$.
While this does not give a complete characterization of the wideband
slope as in \cite{ShenAHM11IT}, it does show that interference alignment
in the low-rate regime does not give the same dramatic gain in performance
as in the large bandwidth and high SNR regimes. We still show that
interference alignment can outperform TDMA, but in line with the outer
bound, not by much.

\section{System Model And Preliminaries}

In \cite{ShenAHM11IT} we derived the following baseband model for
the interference channel (in a line-of-sight model): 
\begin{eqnarray*}
y_{j}[n] & = & C_{jj}x_{j}[n]+\sum_{i\neq j}C_{ji}\tilde{x}_{i}[n-n_{ji}]+z_{j}[n]
\end{eqnarray*}
where 
\begin{eqnarray}
\tilde{x}_{i}[n] & = & \sum_{m=-\infty}^{\infty}x_{i}[m]\sinc(n-m+\delta_{ji}).\label{tildex.eq}
\end{eqnarray}
and
\begin{eqnarray}
n_{ji} & = & \left\lfloor \tau_{ji}B+{\textstyle \frac{1}{2}}\right\rfloor \\
\delta_{ji} & = & \tau_{ji}B-\left\lfloor \tau_{ji}B+{\textstyle \frac{1}{2}}\right\rfloor \label{fracdelay.eq}
\end{eqnarray}
are the symbol and fractional delays, respectively. It was these delays
that allowed interference alignment in \cite{ShenAHM11IT} as $B\to\infty$.

In the present paper we keep $B$ fixed; we will further assume that
$B$ is so small that the delays are insignificant, $n_{ji}=0,\delta_{ji}\approx0$,
and we therefore arrive at the usual model for the interference channel,

\begin{eqnarray}
y_{j}[n] & = & C_{jj}x_{i}[n]+\sum_{i\neq j}C_{ji}x_{i}[n]+Z_{j}[n],\label{eq:scalar model}
\end{eqnarray}
where $C_{ji}$ is a complex scalar and the noise $Z_{j}$ is i.i.d.
(independent, identically distributed) circularly symmetric complex
random variable with distribution $\mathcal{CN}\left(0,\, BN_{0}\right)$;
since $B$ does not play any role in the rest of the paper we will
put $B=1$ and omit it from future formulas. Notice that the model
(\ref{eq:scalar model}) is valid also for a non line-of-sight model,
as long as delays along all paths are insignificant.

\subsection{\label{sub:Circularly-Asymmetric-Signaling}Circularly Asymmetric
Signaling}

To characterize the Shannon capacity region of the model (\ref{eq:scalar model}),
most research restricts the inputs to be circularly symmetric, i.e.,
the the real part of the input $\mathrm{Re}\left\{ x_{j}\right\} $
and the imaginary part of the input $\mathrm{Im}\left\{ x_{j}\right\} $
are i.i.d.. However, \cite{CadJafWan09CoRR} shows that circularly
asymmetric signaling achieves higher degree of freedom in the high-SNR
regime. Although the specific interference alignment technique they
proposed is not applicable to the low-$\snr$ regime, that work still
has inspired our interference alignment for the low-SNR regime. In
section \ref{sec:Sum-Slope-Achievable}, we will see that circularly
asymmetric signaling indeed benefits system performance. 

In circularly asymmetric signaling, the transmitters are allowed to
allocate power on real and imaginary dimensions, and the real part
of the input $\mathrm{Re}\left\{ x_{j}\right\} $ is allowed to be
correlated with the imaginary part of the input $\mathrm{Im}\left\{ x_{j}\right\} $,
while in circularly symmetric signaling, $\mathrm{Re}\left\{ x_{j}\right\} $
and $\mathrm{Im}\left\{ x_{j}\right\} $ are required to be i.i.d..
To characterize such transmission schemes, it is more convenient to
consider the scalar complex channel as a two-dimensional vector real
channel. 

Following \cite{CadJafWan09CoRR}, we extend (\ref{eq:scalar model})
into an equivalent two-dimensional real channel,
\begin{eqnarray}
\underline{Y}_{j} & = & \left|C_{jj}\right|\underline{X}_{j}+\sum_{i=1,i\neq j}^{K}\left|C_{ji}\right|\mathbf{U}_{ji}\underline{X}_{i}+\underline{Z}_{j}\label{eq:eq vector}
\end{eqnarray}
where $\mathbf{U}_{ji}\triangleq\left(\begin{array}{cc}
\cos\left(\phi_{ji}\right) & -\sin\left(\phi_{ji}\right)\\
\sin\left(\phi_{ji}\right) & \cos\left(\phi_{ji}\right)
\end{array}\right)$ is the rotation matrix with angle $\phi_{ji}$, and the $2\times1$
vector white Gaussian noise is $\underline{Z}_{j}\sim\mathcal{N}\left(0,\,\frac{N_{0}}{2}\mathbf{I}_{2\times2}\right)$.
Notice that we let receiver $j$ be phase-synchronized with the received
$x_{j}$ so that $\phi_{jj}=0$.  Without without loss of generality,
we can assume $N_{0}=1$ whenever convenient.

The input signal $\underline{X}_{j}$ is related to the scalar complex
model by: $\underline{X}_{j}=\left(\begin{array}{c}
\mathrm{Re}\left\{ x_{j}\right\} \\
\mathrm{Im}\left\{ x_{j}\right\} 
\end{array}\right)$. We assume that an $\left(2^{nR_{j}},\, n\right)$ code is used at
receiver $j$, for $j=1,\cdots,\, K$. At the transmitter $j$, the
input message $W_{j}$ is drawn uniformly randomly from the index
set $\left\{ 1,\,\cdots,\,2^{nR_{j}}\right\} $, and a deterministic
function yields the length $n$ transmitted codeword $\underline{X}_{j}^{n}\left(W_{j}\right)$.
The codebook of user $j$ is composed by the set of codewords $\underline{X}_{j}^{n}\left(1\right),\,\cdots,\,\underline{X}_{j}^{n}\left(2^{nR_{j}}\right)$.
We require each user to satisfy power constraint $\nicefrac{P_{j}}{B}$
per second per Hz. Recall that we may assume $B=1$. Denote the $ith$
entry of $\underline{X}_{j}^{n}$ by $\underline{X}_{j}^{\left(i\right)}$.
Therefore the input must satisfy constraint 
\begin{equation}
\frac{1}{n}\sum_{i=1}^{n}E\left[\underline{X}_{j}^{\left(i\right)}\left(\underline{X}_{j}^{\left(i\right)}\right)^{T}\right]\preceq\mathbf{V}_{j},\label{eq:power constraint}
\end{equation}
where $\trace\left(\mathbf{V}_{j}\right)=P_{j}$, $j=1,\cdots,\, K$.
For any two given matrices $\mathbf{A}$ and $\mathbf{B}$, the notation
$\mathbf{A}\preceq\mathbf{B}$ means that the matrix $\mathbf{B}-\mathbf{A}$
is positive semi-definite. Notice that given the assumption $B=N_{0}=1$,
we have 
\begin{eqnarray}
\snr_{j} & = & \frac{P_{j}}{BN_{0}}\nonumber \\
 & = & P_{j}\label{eq:snr}
\end{eqnarray}

Corresponding to the $\underline{X}_{j}^{n}\left(W_{j}\right)$ codebook
, we also define four Gaussian random variables $\underline{X}_{jG}^{\prime}$,
$\underline{X}_{jG}$, $\underline{Y}_{jG}^{\prime}$, and $\underline{Y}_{jG}$
as follows for later use. Let $\underline{X}_{jG}^{\prime}$ be i.i.d.
vector Gaussian random variable, $\underline{X}_{jG}^{\prime}\sim\mathcal{N}\left(0,\,\mathbf{V}_{j}^{\prime}\right)$,
where $\mathbf{V}_{j}^{\prime}=\frac{1}{N}\sum_{n=1}^{N}\underline{X}_{j}^{n}\left(\underline{X}_{j}^{n}\right)^{\mathrm{H}}$,
$\mathbf{V}_{j}^{\prime}\preceq\mathbf{V}_{j}$ given power constraint
(\ref{eq:power constraint}). Let $\underline{X}_{jG}$ be i.i.d.
vector Gaussian random variable, $\underline{X}_{jG}\sim\mathcal{N}\left(0,\,\mathbf{V}_{j}\right)$.
$\underline{Y}_{jG}^{\prime}$ and $\underline{Y}_{jG}$ are defined
as
\begin{eqnarray*}
\underline{Y}_{jG}^{\prime} & = & \left|C_{jj}\right|\underline{X}_{jG}^{\prime}+\sum_{i=1,i\neq j}^{K}\left|C_{ji}\right|\mathbf{U}_{ji}\underline{X}_{iG}^{\prime}+\underline{Z}_{j}\\
\underline{Y}_{jG} & = & \left|C_{jj}\right|\underline{X}_{jG}+\sum_{i=1,i\neq j}^{K}\left|C_{ji}\right|\mathbf{U}_{ji}\underline{X}_{iG}+\underline{Z}_{j}
\end{eqnarray*}

\subsection{Performance Criterion And Performance Measures}

For more than two users it is complicated to compare complete slope
regions, and we are therefore looking at a single quantity--the sum
slope $\slope_{0}$, to characterize performance. The formal definitions
are as follows.
\begin{defn}[Sum slope]
 $\slope_{0}$ is defined as the first-order slope of the $R_{sum}\left(\ebno_{sum}\right)$
curve, where $R_{sum}\triangleq\sum_{j=1}^{K}R_{j}$ and $\ebno_{sum}\triangleq\frac{\sum_{j=1}^{K}P_{j}}{N_{0}B\sum_{j=1}^{K}R_{j}}$.
It characterizes the wideband slope of $R_{sum}$ as $\ebno_{sum}$
approaches its minimum value $\left.\ebno\right|_{\min}$: 
\begin{eqnarray}
\left.\ebno\right|_{\min} & = & \lim_{P_{sum}\downarrow0}\frac{\sum_{j=1}^{K}P_{j}}{\sum_{j=1}^{K}R_{j}\cdot N_{0}B}\label{eq:ebnomin0}\\
S_{0} & \triangleq & \lim_{\ebno_{sum}\downarrow\ebnomin}\frac{R_{sum}\left(\ebno\right)10\log_{10}2}{10\log_{10}\ebno_{sum}-10\log_{10}\ebnomin}\label{eq:slope0}
\end{eqnarray}
Denote the sum power constraint by $P_{sum}=\sum_{j=1}^{K}P_{j}$.
Under the assumption that $N_{0}B=1$, $\left.\ebno\right|_{\min}$
and $S_{0}$ can be obtained from the first and second order derivatives
of $R_{sum}\left(P_{sum}\right)$: 
\begin{eqnarray}
\left.\ebno\right|_{\min} & = & \frac{\log_{e}2}{\dot{R}_{sum}\left(0\right)};\label{eq:ebnomin0-1}\\
S_{0} & = & -\frac{2\left(\dot{R}_{sum}\left(0\right)\right)^{2}}{\ddot{R}_{sum}\left(0\right)}.\label{eq:slope0-1}
\end{eqnarray}

\end{defn}
Notice that constraints on $P_{j}$ or $R_{j}$ are required for a
well-posed problem; otherwise the best low-SNR performance is achieved
by allocating all power to the user with largest direct link gain
so that $\left.\ebno\right|_{\min}$ is minimized. Such a solution
is just a single user solution and gives no insight into the interference
channel. To fix this insufficiency while keeping our problem relatively
simple to analyze, we require the interference channel to work under
the \emph{equal-power constraint, }which is defined as
\begin{defn}
\emph{\label{Equal-power-constraint}Equal power constraint} is the
case where the sum rate $R_{sum}$ is maximized under the constraint
$P_{1}=P_{2}=\cdots=P_{K}$. 
\end{defn}
Given (\ref{eq:slope def}), we can see that \emph{if two systems
achieve equal $\ebnomin$ value}, the $\ebno$ value of the system
with higher wideband slope approaches its minimum value faster, and
the system is therefore more spectrally efficient. On the other hand,
we should notice that the priority in the low-$\snr$ regime is to
minimize $\ebnomin$. Based on this observation, we make the following
statement:
\begin{rem}
\label{rem:To-make-fair}To make fair comparison of the wideband slopes
between different systems, they must have equal $\ebnomin$ in the
first place. 
\end{rem}
The results in \cite{CaiTunVer04IT} reveal that the optimal achievable
minimum energy per bit $\ebnomin$ of an interference channel is equal
to that of its corresponding interference-free channel. The first-order
optimality criterion under the equal power constraint is stated in
the following lemma. 
\begin{lem}
\label{lem: correct ebnomin}The optimal minimum energy per bit of
the interference channel defined by (\ref{eq:scalar model}) is 
\begin{eqnarray}
\frac{E_{b}}{N_{0}}_{\mathrm{min}} & = & \frac{K\log_{e}2}{\sum_{j=1}^{K}\left|C_{jj}\right|^{2}}\label{eq:ebnomin equal power}
\end{eqnarray}
under the equal power constraint.
\end{lem}
\emph{Given Remark \ref{rem:To-make-fair}, any achievable scheme
or capacity outer bound gives valid bound on the sum slope only if
it has correct $\ebnomin$ values, stated in Theorem \ref{lem: correct ebnomin}.}

For performance measure we use
\begin{eqnarray*}
\Delta\mathcal{S}_{0} & = & \frac{\mathcal{S}_{0}}{\mathcal{S}_{0,\mbox{no interference}}}.
\end{eqnarray*}
The quantity $\mathcal{S}_{0,\mbox{no inteference}}$ is the wideband
slope of the corresponding interference-free channel: 
\begin{eqnarray*}
R_{j} & = & \log\left(1+\left|C_{jj}\right|^{2}P_{j}\right).
\end{eqnarray*}
 We can interpret $\Delta\mathcal{S}_{0}$ as the loss in wideband
slope due to interference. 

Under the equal power constraint, $\mathcal{S}_{0,\mathrm{no\, interference}}$
the sum slope of the interference-free channel, and $\mathcal{S}_{0,TDMA}$
and $\mathcal{S}_{0,TIN}$ the sum slope achieved by TDMA and treating
interference as noise (TIN) respectively, are listed as follows for
comparison purposes; they can be obtained directly obtained from (\ref{eq:ebnomin calc}-\ref{eq:slope calc})
\begin{eqnarray}
\begin{array}{rcll}
\mathcal{S}_{0,\mathrm{no\, interference}} & = & 2{\displaystyle \frac{\left(\sum_{j}\left|C_{jj}\right|^{2}\right)^{2}}{\sum_{j}\left|C_{jj}\right|^{4}}}\end{array}\label{eq:comp ep}
\end{eqnarray}
The $R_{sum}\left(P_{sum}\right)$ achieved by TIN is 
\begin{eqnarray*}
R_{sum}\left(P_{sum}\right) & = & \sum_{j=1}^{K}\log\left(1+\frac{\left|C_{jj}\right|^{2}P_{sum}}{K+P_{sum}\sum_{i\neq j}\left|C_{ji}\right|^{2}}\right),
\end{eqnarray*}
which gives 
\begin{eqnarray}
\slope_{0,TIN} & = & 2{\displaystyle \frac{\left(\sum_{j}\left|C_{jj}\right|^{2}\right)^{2}}{\sum_{j=1}^{K}\left(\left|C_{jj}\right|^{4}+2\sum_{i\neq j}\left|C_{ji}\right|^{2}\left|C_{jj}\right|^{2}\right)}};\label{eq:sum slope TIN}\\
\Delta\slope_{0} & = & \frac{\sum_{j=1}^{K}\left|C_{jj}\right|^{4}}{\sum_{j=1}^{K}\left(\left|C_{jj}\right|^{4}+2\sum_{i\neq j}\left|C_{ji}\right|^{2}\left|C_{jj}\right|^{2}\right)}
\end{eqnarray}
The $R_{sum}\left(P_{sum}\right)$ achieved by TDMA is 
\begin{eqnarray*}
R_{sum}\left(P_{sum}\right) & = & \frac{1}{K}\sum_{j=1}^{K}\log\left(1+\left|C_{jj}\right|^{2}P_{sum}\right),
\end{eqnarray*}
which gives 
\begin{eqnarray}
\slope_{0} & = & \frac{2}{K}{\displaystyle \frac{\left(\sum_{j}\left|C_{jj}\right|^{2}\right)^{2}}{\sum_{j}\left|C_{jj}\right|^{4}}}\label{eq:sum slope TDMA}\\
\Delta\slope_{0} & = & \frac{1}{K}\label{eq:sum slope TDMA-1}
\end{eqnarray}

\section{\label{sec:Generalized-Z-Channel-Outer}Generalized Z-Channel Outer
Bound}

In this section, we develop a new outer bound on the wideband slope
for a set of the 2-dimensional vector channels defined by (\ref{eq:eq vector}),
under the equal power constraint. The outer bound is specific to the
low-rate regime.

The outer bound is derived from the sum Shannon capacity of a type
of generalized Z-channel, which is constructed by elimination of a
subset of the interference links. In Section \ref{sub:Z sum cap},
we show that for a subset of channels $\mathcal{C}$, the optimal
sum capacity of their corresponding Z-channels can be achieved by
i.i.d. 2-dimensional vector Gaussian inputs. Further, assuming that
channel coefficients $C_{ji}$ is drawn from i.i.d. continuous distribution,
the set $\mathcal{C}$ has non-zero probability. In Section \ref{sub:Sum-Slope-Outer},
the Z-channel outer bound is used to derive an outer bound on the
wideband slope.

\subsection{\label{sub:Z sum cap}Generalized Z-Channel And Its Sum Capacity}

We define the generalized Z-channel corresponding to the interference
channel (\ref{eq:eq vector}) as 
\begin{eqnarray}
\underline{\hat{Y}}_{j} & = & \left|C_{jj}\right|\underline{X}_{j}+\sum_{i=j+1}^{K}\left|C_{ji}\right|\mathbf{U}_{ji}\underline{X}_{i}+\underline{Z}_{j}.\label{eq:k-user z-channel}
\end{eqnarray}
\begin{figure}[H]
\begin{centering}
\includegraphics[width=3.5in]{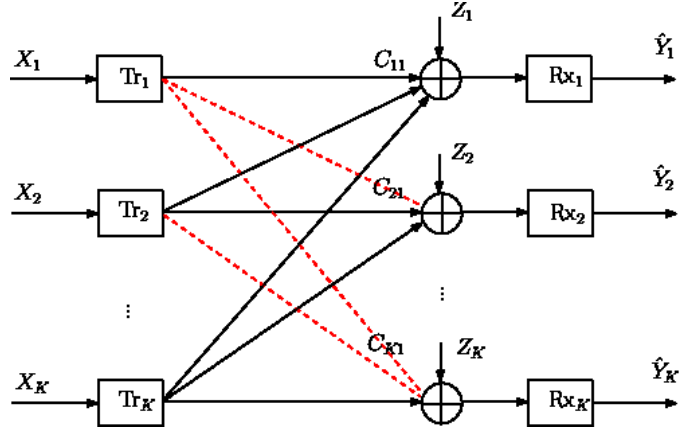}
\par\end{centering}

\caption{{\small Generalized Z-channel}}
\end{figure}
Eliminating a subset of interference links will not reduce channel
capacity and therefore, the sum capacity outer bound for the generalized
Z-channel is also a sum capacity outer bound for the interference
channel. 

To derive the Z-channel sum capacity, we provide receiver $j$, $j=2,\cdots,\, K$
with side information $\mathbf{\underline{S}}_{j}^{n}=\left(\underline{S}_{j1}^{n},\cdots,\,\underline{S}_{j\left(j-1\right)}^{n}\right)^{T}$,
where 

\begin{eqnarray}
\underline{S}_{jp}^{n} & = & \left|C_{pj}\right|\mathbf{U}_{pj}\underline{X}_{j}^{n}+\sum_{i=j+1}^{K}\left|C_{pi}\right|\mathbf{U}_{pi}\underline{X}_{i}^{n}+\underline{W}_{jp}^{n}\label{eq:side info}
\end{eqnarray}
$p=1,\cdots,\, j-1$. . The entries in the length $n$ noise vector
$\underline{W}_{jp}^{n}$ are i.i.d $2\times1$ vector Gaussian noise
with the same marginal distribution as $Z_{j}$. Further, they satisfy
the following properties
\begin{itemize}
\item $\underline{W}_{j\left(j-1\right)}^{n},\cdots,\,\underline{W}_{j1}^{n}$
are independent of all input length $n$ codewords $\underline{X}_{i}^{n}$,
$i=1,\cdots,\, K$;
\item $\left(\underline{Z}_{j},\underline{W}_{j\left(j-1\right)},\cdots,\,\underline{W}_{j1}\right)$
are jointly Gaussian random variables, with zero mean and covariance
matrix
\end{itemize}
\begin{equation}
\mathbf{K}_{S_{j}}=\left(\begin{array}{ccccc}
\mathbf{I} & \mathbf{A}_{j\left(j-1\right)} & \cdots & \mathbf{A}_{j1} & \mathbf{A}_{j1}\\
\mathbf{A}_{j\left(j-1\right)}^{T} & \mathbf{I} & \mathbf{A}_{\left(j-1\right)\left(j-2\right)} & \cdots & \mathbf{A}_{\left(j-1\right)1}\\
\vdots &  & \ddots & \ddots & \vdots\\
\mathbf{A}_{j1}^{T} &  &  & \mathbf{I} & \mathbf{A}_{21}\\
\mathbf{A}_{j1}^{T} & \mathbf{A}_{\left(j-1\right)1}^{T} & \cdots & \mathbf{A}_{21}^{T} & \mathbf{I}
\end{array}\right)\label{eq:w2}
\end{equation}

To guarantee such multivariate Gaussian random variable exists, $\mathbf{A}_{jk}$
should be chosen such that for all $j=1,\cdots,\, K$ 
\begin{eqnarray}
\mathbf{K}_{S_{j}} & \succeq & 0\label{eq:cova}
\end{eqnarray}

We emphasize the following property of $\mathbf{K}_{S_{j}}$, which
will play a key role in the proof of the main result.
\begin{lem}
\label{lem:marginal_1}The distributions of $\left.\underline{S}_{\left(j-1\right)p}^{n}\right|\underline{S}_{\left(j-1\right)\left(p-1\right)}^{n},\cdots,\,\underline{S}_{\left(j-1\right)1}^{n},\underline{X}_{\left(j-1\right)}^{n}$
and $\left.\underline{S}_{jp}^{n}\right|\underline{S}_{j\left(p-1\right)}^{n},\cdots,\,\underline{S}_{j1}^{n}$
are equal. 
\end{lem}
Proof of Lemma \ref{lem:marginal_1} is in Appendix \ref{sec:Proof Lemma marginal}. 
\begin{lem}
\label{lem:marginal 2}The distributions of $\left.\underline{\hat{Y}}_{j-1}^{n}\right|\underline{S}_{\left(j-1\right)\left(j-2\right)}^{n},\cdots,\,\underline{S}_{\left(j-1\right)1}^{n},\underline{X}_{\left(j-1\right)}^{n}$
and $\left.\underline{S}_{j\left(j-1\right)}^{n}\right|\underline{S}_{j\left(j-2\right)}^{n},\cdots,\,\underline{S}_{j1}^{n}$
are equal. 
\end{lem}
The proof of Lemma \ref{lem:marginal 2} is almost identical to the
proof of Lemma \ref{lem:marginal_1} and will therefore be omitted.

Define the average covariance matrix of the input at transmitter as
\[
\tilde{\mathbf{V}}_{j}\triangleq\frac{1}{n}\sum_{i=1}^{n}\mathrm{E}\left[\underline{X}_{j}^{\left(i\right)}\left(\underline{X}_{j}^{\left(i\right)}\right)^{T}\right]
\]
 for any length $n$ input sequence $\underline{X}_{j}^{n}$. It must
satisfy the power constraint defined in (\ref{eq:power constraint}),
i.e., $\tilde{\mathbf{V}}_{j}\preceq\mathbf{V}_{j}$. The next lemma
states how to choose $\mathbf{A}_{jk}$.
\begin{lem}
\label{lem:side-info}Let $\mathbf{A}_{jp}$, $j=2,\cdots,\, K$ and
$p=1,\cdots,\, j-1$ be 

\begin{eqnarray}
\mathbf{A}_{jp} & = & \frac{\left|C_{pj}\right|^{2}}{\left|C_{jj}\right|^{2}}\mathbf{U}\left(-\phi_{pj}\right)\nonumber \\
 &  & +\frac{\left|C_{pj}\right|^{2}}{\left|C_{jj}\right|^{2}}\sum_{i=j+1}^{K}\left|C_{ji}\right|^{2}\mathbf{U}\left(\phi_{ji}\right)\mathbf{V}_{i}\mathbf{U}\left(-\phi_{pj}-\phi_{ji}\right)\nonumber \\
 &  & -\sum_{i=j+1}^{K}\left|C_{ji}\right|^{2}\left|C_{pi}\right|^{2}\mathbf{U}\left(\phi_{ji}\right)\mathbf{V}_{i}\mathbf{U}\left(-\phi_{pi}\right)\label{eq:side-info cov}
\end{eqnarray}

If $\mathbf{A}_{jp}$ defined by (\ref{eq:side-info cov}) satisfy
$\mathbf{K}_{S_{j}}\succeq0$, then 

\begin{eqnarray}
 & \underline{X}_{jG}\rightarrow & \underline{\hat{Y}}_{jG}\rightarrow\left(\underline{S}_{j1G},\cdots,\,\underline{S}_{j\left(j-1\right)G}\right)^{T}\label{eq:markov 1}
\end{eqnarray}

forms a Markov chain for all $j=2,\cdots,\, K$. 
\end{lem}
Here $\underline{X}_{jG}$ and $\underline{\hat{Y}}_{jG}$ are defined
in section \ref{sub:Circularly-Asymmetric-Signaling}; the proof of
Lemma \ref{lem:side-info} is in Appendix \ref{sec:Proof lemma side-info}. 

For a channel realization, denote its channel coefficients by $\underline{C}\triangleq\left\{ C_{ji};\, i,\, j=1,\cdots,\, K\right\} $.
In the following lemma, we state a sufficient condition on $\underline{C}$
so that $\mathbf{K}_{S_{j}}\succeq0$ if $\mathbf{A}_{jp}$ is chosen
according to (\ref{eq:side-info cov}). 
\begin{lem}
\label{lem:sufficient} For any $0<\alpha<1$ there exist some $\epsilon_{\alpha},\epsilon_{\alpha}^{\prime}>0$
and $\epsilon_{\alpha}^{\prime\prime}(\underline{C})>0$ so that if
\begin{eqnarray}
\underline{C} & \in & \mathcal{C}_{\alpha}\triangleq\left\{ C_{ij}:\,\left|\frac{\left|C_{ij}\right|^{2}}{\left|C_{jj}\right|^{2}}-\alpha\right|<\epsilon_{\alpha},\right.\nonumber \\
 &  & \left.\left|\phi_{ji}\right|<\epsilon_{\alpha}^{\prime}\right\} \label{eq:cond h}\\
P_{j} & < & \epsilon_{\alpha}''(\underline{C})\label{cond1-1}
\end{eqnarray}
then $\mathbf{K}_{S_{j}}\succeq0$ for $\mathbf{A}_{jp}$ chosen according
to (\ref{eq:side-info cov}).
\end{lem}
Proof of Lemma \ref{lem:sufficient} is in Appendix \ref{sec:Proof-of-Lemma sufficient}

Our main result of this section is stated in the following theorem. 
\begin{thm}
\label{thm:K-user sum rate}For every interference channel realization
$\underline{C}\in\mathcal{C}=\bigcup_{\alpha\in\left(0,1\right)}\mathcal{C}_{\alpha}$
defined by (\ref{eq:cond h}) there exists an $\epsilon_{\alpha}''(\underline{C})>0$
so that if $P_{j}<\epsilon_{\alpha}''(\underline{C})$ the sum capacity
of its corresponding Z-channel is given by

\begin{eqnarray}
\sum_{j=1}^{K}R_{j}\leq C_{\mathrm{sum}} & = & \max_{\begin{array}{c}
\trace\left(\mathbf{V}_{j}\right)\leq P_{j}\\
\mathbf{V}_{j}\succeq\mathbf{0},\, j=1,\cdots,K
\end{array}}\sum_{j=1}^{K}I\left(\underline{X}_{jG};\underline{\hat{Y}}_{jG}\right)\label{eq:K-user sum rate bound}\\
 & = & \max_{\begin{array}{c}
\trace\left(\mathbf{V}_{j}\right)\leq P_{j}\\
\mathbf{V}_{j}\succeq\mathbf{0},\, j=1,\cdots,K
\end{array}}\sum_{j=1}^{K}\log\left|\left(\mathbf{I}+\sum_{i=j}^{K}\left|C_{ji}\right|^{2}\mathbf{V}_{i}\right)\left(\mathbf{I}+\sum_{i=j+1}^{K}\left|C_{ji}\right|^{2}\mathbf{V}_{i}\right)^{-1}\right|
\end{eqnarray}
Because the sum capacity of the interference channel is outer bounded
by the sum capacity of the generalized Z-channel, (\ref{eq:K-user sum rate bound})
is an outer bound for the sum capacity of the interference channel.
\end{thm}
Proof of Theorem \ref{thm:K-user sum rate} is in Appendix \ref{sec:z-channel sum rate}.

Note that the bound in Theorem \ref{thm:K-user sum rate} is valid
for $P_{j}<\epsilon_{\alpha}^{\prime\prime}(\underline{C})$, and
it therefore bounds the actual capacity for suitably low SNR. However,
we will mainly use it to bound the wideband slope, a weaker result.

\subsection{\label{sub:Sum-Slope-Outer}Sum Slope Outer Bound for the Interference
Channel}

Given the capacity in Theorem \ref{thm:K-user sum rate}, we have
following result on the low-rate performance of the interference channel. 
\begin{thm}
\label{thm:sum slope}For the interference channel (\ref{eq:scalar model}),
the sum capacity is outer bounded by (\ref{eq:K-user sum rate bound})
for low $\snr$. Under the equal power constraint, the minimum energy
per bit of this upper bound satisfy the requirement imposed by Remark
\ref{rem:To-make-fair}, which is 
\begin{eqnarray}
\ebnomin & = & \frac{K\log2}{\sum_{j=1}^{K}\left|C_{jj}\right|^{2}}\label{eq:ebnomin}
\end{eqnarray}
For channel realizations $\underline{C}\in\mathcal{C}=\bigcup_{\alpha\in\left(0,1\right)}\mathcal{C}_{\alpha}$
defined as (\ref{eq:cond h}) it therefore gives the following valid
upper bound on the sum slope: 
\begin{eqnarray}
\slope_{0} & \leq & \left(\sum_{j=1}^{K}\left|C_{jj}\right|^{2}\right)^{2}\label{eq:slope detail-1}\\
 &  & \times\max_{\begin{array}{c}
\trace\left(\hat{\mathbf{V}}_{j}\right)\leq1\\
\hat{\mathbf{V}}_{j}\succeq\mathbf{0}
\end{array}}\left(\sum_{j=1}^{K}\left|C_{jj}\right|^{4}\trace\left(\hat{\mathbf{V}}_{j}^{2}\right)\right.\\
 &  & \left.+2\sum_{j=1}^{K-1}\sum_{i=j+1}^{K}\left|C_{jj}\right|^{2}\left|C_{ji}\right|^{2}\trace\left(\hat{\mathbf{V}}_{j}\mathbf{U}_{ji}\hat{\mathbf{V}}_{i}\mathbf{U}_{ji}^{\dagger}\right)\right)^{-1}
\end{eqnarray}

\end{thm}
Proof of Theorem \ref{thm:sum slope} is in Appendix \ref{sec:Proof-of-sum slope outer}.
\begin{thm}
\label{cor:TDMA0}For the symmetric channel where $C_{jj}=1$, $C_{ji}=\alpha\in(0,1)$
, the sum slope is bounded by
\begin{eqnarray*}
\slope_{0} & \leq & \frac{2K}{\alpha K+\left(1-\alpha\right)}
\end{eqnarray*}

\end{thm}
Proof of Theorem \ref{cor:TDMA0} is in Appendix \ref{sec: proof TDMA}.

As discussed in the introduction, the wideband slope in the point
$\underline{C}=\mathbf{1}$ is $\frac{2}{K}$ per user, achievable
by TDMA. Theorem \ref{cor:TDMA0} shows that the point $\underline{C}=\mathbf{1}$
is not exceptional in the low-rate regime: for $\alpha$ close to
1 (from below) the channel with $C_{jj}=1,C_{ji}=\alpha$ has slope
close to $\frac{2}{K}$. However, the set of channels $C_{jj}=1,C_{ji}=\alpha$
still has Lebesgue measure zero, i.e., if the channel coefficients
are drawn from a continuous distribution, this set has probability
zero. The main result of the paper is the following theorem that shows
that the set of channels with slope close to $\frac{2}{K}$ can be
be extended to a set of non-zero measure.
\begin{thm}
\label{cor: TDMA}For all $\sigma>0$, there exists an open set $\tilde{\mathcal{C}}_{\sigma}\subset\mathbb{C}^{K(K-1)}$
with $\mathbf{1}\in\mbox{cl}\left(\tilde{\mathcal{C}}_{\sigma}\right)$,
so that for $\underline{C}\in\tilde{\mathcal{C}}_{\sigma}$
\begin{eqnarray}
\slope_{0} & \leq & 2+\sigma,\label{eq:almost TDMA}
\end{eqnarray}
If the magnitude and phase of the channel coefficients are drawn from
continuous random distribution, $Pr\left(\tilde{\mathcal{C}}_{\sigma}\right)>0$. 

And as $\sigma\rightarrow0$, 
\begin{eqnarray*}
\lim_{\sigma\rightarrow0}\Delta\slope_{0} & = & \frac{1}{K}
\end{eqnarray*}
Because $\Delta\slope_{0}$ achieved by TDMA is $\frac{1}{K}$, when
$\sigma$ is small, TDMA transmission scheme is almost optimal for
channels in $\tilde{\mathcal{C}}_{\sigma}$. 
\end{thm}
Proof of Theorem \ref{cor: TDMA} is in Appendix \ref{sec: proof TDMA-1}.

\section{\label{sec:Sum-Slope-Achievable}Sum Slope Achievable Scheme}

In the previous section, we have shown that there exist a set of channels
$\mathcal{C}_{\sigma}$, $Pr\left(\tilde{\mathcal{C}}_{\sigma}\right)>0$,
for which TDMA is almost optimal. However, we also notice that the
probability that a channel realization is not in $\mathcal{C}_{\sigma}$
is likewise greater than zero. Therefore, it is natural to ask the
question: for channels not in $\tilde{\mathcal{C}}_{\sigma}$, can
we find achievable schemes better than TDMA or Treating Interference
as Noise (TIN)? 

In section \ref{sub:One-Dimensional-Gaussian-Signali}, we propose
a circularly asymmetric transmission scheme and analyze its theoretical
performance. Simulation results are shown in section \ref{sub:Simulation-Results-and}.
We will also discuss possible improvements of this scheme.

\subsection{\label{sub:One-Dimensional-Gaussian-Signali}One-Dimensional Gaussian
Signaling}

In this section, we use the complex scalar channel model defined in
(\ref{eq:scalar model}). We define a one-dimensional Gaussian signaling
transmission scheme and analyze its performance. The idea is to align
interference as much as possible.
\begin{defn}
\label{def: One-dim-sig}One-dimensional Gaussian signaling transmission
scheme
\begin{itemize}
\item At transmitter $j$, let input sequence be $x_{j}\left[n\right]=w_{j}\left[n\right]e^{j\theta_{j}}$,
where $w_{j}\left[n\right]$ is drawn from i.i.d real Gaussian random
variable with distribution $\mathcal{N}\left(0,\,\snr_{j}\right)$,
and the phase $\theta_{j}$ is a prior chosen design parameter, unchanged
for all $n$ during the transmission.
\item At receiver $j$, interference is treated as noise.
\end{itemize}
\end{defn}
We call this one-dimensional because every transmitter only transmits
along $e^{j\theta_{j}},$ therefore only one dimension is used out
of the two-dimensional signal space. 

Our objective is to find the set of phases $\underline{\theta}=\left\{ \theta_{1},\cdots,\,\theta_{K}\right\} $
that maximize the achievable wideband slope $\slope_{0}$. 

The achievable $\slope_{0}$ for any $\underline{\theta}$ is stated
in the next lemma. For computational convenience, we return to the
equivalent two-dimensional real channel model. In the equivalent 2-dimensional
real channel model, the input $\underline{X}_{j}$ has covariance
matrixw
\begin{eqnarray*}
\mathbf{V}_{j} & = & P_{j}\left(\begin{array}{cc}
\cos^{2}\theta & \frac{\sin2\theta}{2}\\
\frac{\sin2\theta}{2} & \sin^{2}\theta
\end{array}\right),
\end{eqnarray*}
$rank\left(\mathbf{V}_{j}\right)=1$. We denote the normalized covariance
matrix by $\hat{\mathbf{V}}_{j}=\frac{\mathbf{V}_{j}}{P_{j}}$. 
\begin{lem}
For the equivalent 2-dimensional real channel model defined by (\ref{eq:eq vector}),
the sum slope achieved by the one-dimensional Gaussian signaling is
\begin{eqnarray}
\slope_{0} & = & \frac{\left(\sum_{j=1}^{K}\left|C_{jj}\right|^{2}\right)^{2}}{\sum_{j=1}^{K}\left|C_{jj}\right|^{4}+\sum_{j=1}^{K}\sum_{i\neq j}^{K}\left|C_{jj}\right|^{2}\left|C_{ji}\right|^{2}+f\left(\underline{\theta}\right)},\label{eq:lem11}
\end{eqnarray}
where 
\begin{eqnarray}
f\left(\underline{\theta}\right) & \triangleq & \sum_{j=1}^{K}\sum_{i\neq j}^{K}\left|C_{jj}\right|^{2}\left|C_{ji}\right|^{2}\cos2\left(\phi_{ji}-\theta_{j}+\theta_{i}\right).\label{eq:optimization problem-1}
\end{eqnarray}
\end{lem}
\begin{IEEEproof}
Treating interference as noise at the receiver, the achievable sum
rate \ref{def: One-dim-sig} is 
\begin{eqnarray}
R_{sum} & =\sum_{j=1}^{K} & \left(\frac{1}{2}\log\left|\mathbf{I}_{2}+\frac{2}{K}P_{sum}\left(\left|C_{jj}\right|^{2}\mathbf{U}\left(\phi_{jj}\right)\hat{\mathbf{V}}_{j}\mathbf{U}_{2}\left(-\phi_{jj}\right)\right.\right.\right.\nonumber \\
 &  & \left.\left.+\sum_{i=1,i\neq j}^{K}\left|C_{ji}\right|^{2}\mathbf{U}\left(\phi_{ji}\right)\hat{\mathbf{V}}_{i}\mathbf{U}\left(-\phi_{ji}\right)\right)\right|\label{eq:rate}\\
 &  & \left.-\frac{1}{2}\log\left|\mathbf{I}_{2}+\frac{2}{K}P_{sum}\sum_{i=1,i\neq j}^{K}\left|C_{ji}\right|^{2}\mathbf{U}\left(\phi_{ji}\right)\hat{\mathbf{V}}_{i}\mathbf{U}\left(-\phi_{ji}\right)\right|\right)
\end{eqnarray}
under the equal power constraint where $\snr_{j}=\frac{\snr_{s}}{K}$.
Combining (\ref{eq:ebnomin calc}), (\ref{eq:slope calc}) and (\ref{eq:rate}),
we have 
\begin{eqnarray}
\dot{R}_{s}\left(0\right) & = & \frac{\sum_{j=1}^{K}\left|C_{jj}\right|^{2}}{K}\label{eq:rsd1}\\
-\ddot{R}_{s}\left(0\right) & = & \frac{2\sum_{j=1}^{K}\left|C_{jj}\right|^{4}}{K^{2}}+\frac{2\sum_{j=1}^{K}\sum_{i\neq j}^{K}\left|C_{jj}\right|^{2}\left|C_{ji}\right|^{2}}{K^{2}}\nonumber \\
 &  & +\frac{2}{K^{2}}\sum_{j=1}^{K}\sum_{i\neq j}^{K}\left(\left|C_{jj}\right|^{2}\left|C_{ji}\right|^{2}\cdot\right.\nonumber \\
 &  & \left.\cos2\left(\phi_{ji}-\theta_{j}+\theta_{i}\right)\right).\label{eq:rsd2}
\end{eqnarray}
Given $\slope_{0}=\frac{2\dot{R}_{s}^{2}\left(0\right)}{-\ddot{R}_{s}\left(0\right)},$
(\ref{eq:lem11}) follows. 
\end{IEEEproof}
Given (\ref{eq:lem11}), maximizing $\slope_{0}$ is equivalent to
finding the set of $\theta_{j}$ that minimizes $f\left(\underline{\theta}\right).$ 

Denote this optimization problem by $P\left(\underline{\theta}\right)$,
which is defined as
\begin{eqnarray*}
 & \min & f\left(\underline{\theta}\right)\\
 & \mathrm{subject\; to} & \theta_{j}\in\left[-\pi,\pi\right].
\end{eqnarray*}
Notice that $\theta_{j}\mod\;2\pi$ will not affect the value of $f\left(\underline{\theta}\right)$.
Given the definition of the objective function in (\ref{eq:optimization problem-1}),
the constraint $\theta_{j}\in\left[-\pi,\pi\right]$ can be discarded.
Therefore, $P\left(\underline{\theta}\right)$ can be solved using
standard numerical methods for unconstrained optimization problems.

\subsection{\label{sub:Simulation-Results-and}Simulation Results and Discussions}

In this section, we simulate the performance of the one-dimensional
signaling scheme in a 10-user interference channel with unit direct
link gains and symmetric weak interference link gains, i.e., $\left|C_{jj}\right|^{2}=1$
and $\left|C_{ji}\right|^{2}=a<1$ for all $i,j=1,\cdots,\,10$; the
phases $\phi_{ji}$ is drawn from $U\left[-\pi,\pi\right]$ in each
channel realization. This performance will be compared with existing
achievable schemes: treating interference as noise and TDMA. 

The simulation results are presented below. We can see that when $\alpha$,
the ratio between the direct link gain and the interference link gain,
is close to 1, then with non-zero probability the one dimensional
Gaussian signaling transmission scheme performs better than TDMA.

\begin{figure}[H]
\begin{centering}
\includegraphics[width=3.5in]{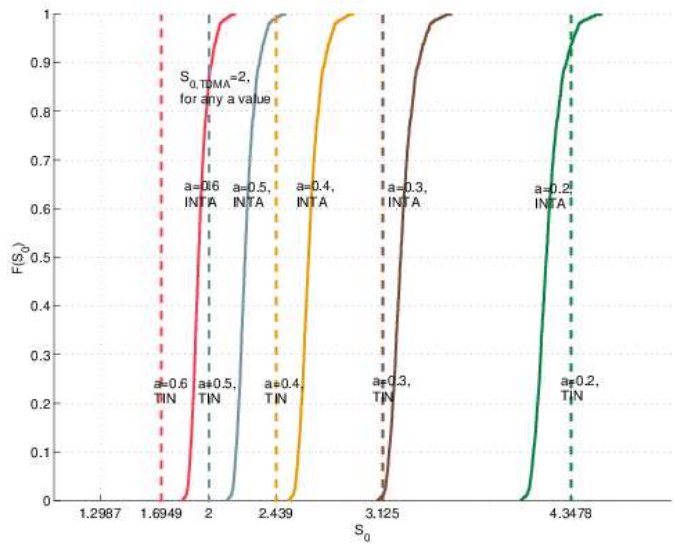}
\par\end{centering}

\caption{\label{fig:cdf}Empirical cumulative distribution functions of $\slope_{0}$
achieved by treating interference as noise (TIN), interference alignment
(INTA) and TDMA under different $a$ values.}
\end{figure}

Fig. \ref{fig:cdf} illustrates the empirical cumulative distribution
functions of the sum slope achieved by the one-dimensional interference
alignment scheme at different $a$ values. For comparison, $\slope_{0}$
achieved by treating interference as noise are also shown, and TDMA
always achieves$\slope_{0}=2$ for all $a$ value.

\begin{figure}[H]
\begin{centering}
\includegraphics[width=3.5in]{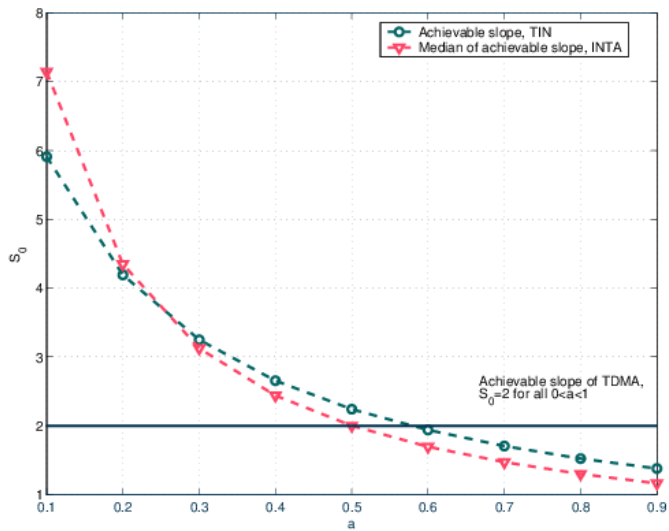}
\par\end{centering}

\caption{\label{fig:single realization}the median value of $\slope_{0}$ achieved
by INTA, and the achievable $\slope_{0}$ of TIN and TDMA as a function
of $a$}
\end{figure}

In Figure \ref{fig:single realization}, we compare the median value
of $\slope_{0}$ achieved by one-dimensional interference alignment
scheme with the performance of treating interference as noise and
TDMA. 

\section{Conclusion}

The main result of this paper can be summarized as follows. In the
low rate regime, the wideband slope is (upper semi-) continuous in
the point $\underline{1}$, the point where all channels are identical,
and where the wideband slope (per user) is $\frac{2}{K}$. This does
not give a full characterization of the wideband slope. However, it
is a stark contrast to the large bandwidth regime \cite{ShenAHM11IT},
where a wideband slope of 1 is achievable almost everywhere, implying
discontinuity in the point $\underline{1}$. It is also a contrast
to the high $\snr$ regime, where $\frac{1}{2}$ DoF per user is achievable
almost everywhere \cite{CadambeJafar07,MotGhaKha09CoRR,WuShamaiVerdu12},
and where the DoF is discontinuous almost everywhere. The results
in \cite{ShenAHM11IT} and \cite{CadambeJafar07,MotGhaKha09CoRR}
were obtained by using interference alignment, and the result in this
paper implies that interference alignment does not give the dramatic
gains in the low rate regime seen elsewhere. Yet, we show that interference
alignment can still give some gain.

One implication of the result is that in networks, as opposed to point-to-point
channels, it is important how the low $\snr$ regime is approached.
This may effect how networks are designed and operates for maximum
energy efficiency.

\bibliographystyle{IEEEtran}
\bibliography{wbslope,Coop06,Nonlinear_programming,Coop03,ahmref2}

\begin{thebibliography}{10}
\providecommand{\url}[1]{#1}
\csname url@samestyle\endcsname
\providecommand{\newblock}{\relax}
\providecommand{\bibinfo}[2]{#2}
\providecommand{\BIBentrySTDinterwordspacing}{\spaceskip=0pt\relax}
\providecommand{\BIBentryALTinterwordstretchfactor}{4}
\providecommand{\BIBentryALTinterwordspacing}{\spaceskip=\fontdimen2\font plus
\BIBentryALTinterwordstretchfactor\fontdimen3\font minus
  \fontdimen4\font\relax}
\providecommand{\BIBforeignlanguage}[2]{{%
\expandafter\ifx\csname l@#1\endcsname\relax
\typeout{** WARNING: IEEEtran.bst: No hyphenation pattern has been}%
\typeout{** loaded for the language `#1'. Using the pattern for}%
\typeout{** the default language instead.}%
\else
\language=\csname l@#1\endcsname
\fi
#2}}
\providecommand{\BIBdecl}{\relax}
\BIBdecl

\bibitem{ShenAHM11IT}
M.~Shen and A.~H{\o}st-Madsen, ``The wideband slope of interference channels:
  The large bandwidth case,'' \emph{IEEE Transactions on Information Theory},
  Submitted, available at http://arxiv.org/abs/1010.5661.

\bibitem{Ver02IT}
S.~Verd{\'u}, ``Spectral efficiency in the wideband regime,'' \emph{IEEE
  Transactions on Information Theory}, vol.~48, no.~6, pp. 1319--1343, 2002.

\bibitem{CadambeJafar07}
V.~Cadambe and S.~Jafar, ``Interference alignment and degrees of freedom of the
  $k$-user interference channel,'' \emph{Information Theory, IEEE Transactions
  on}, vol.~54, no.~8, pp. 3425 --3441, aug. 2008.

\bibitem{MotGhaKha09CoRR}
A.~S. Motahari, S.~O. Gharan, and A.~K. Khandani, ``Real interference alignment
  with real numbers,'' \emph{IEEE Transactions on Information Theory},
  Submitted, available online at http://arxiv.org/abs/0908.1208.

\bibitem{WuShamaiVerdu12}
Y.~Wu, S.~Shamai, and S.~Verdu, ``Degrees of freedom of the interference
  channel: A general formula,'' in \emph{Information Theory Proceedings (ISIT),
  2011 IEEE International Symposium on}, 31 2011-aug. 5 2011, pp. 1362 --1366.

\bibitem{CadJafWan09CoRR}
V.~R. Cadambe, S.~A. Jafar, and C.~Wang, ``Interference alignment with
  asymmetric complex signaling - settling the {H}ost-{M}adsen-{N}osratinia
  conjecture,'' \emph{CoRR}, vol. abs/0904.0274, 2009.

\bibitem{CaiTunVer04IT}
G.~Caire, D.~Tuninetti, and S.~Verd{\'u}, ``Suboptimality of {TDMA} in the
  low-power regime,'' \emph{IEEE Transactions on Information Theory}, vol.~50,
  no.~4, pp. 608--620, 2004.

\bibitem{ShangChenKramerIT10}
X.~Shang, B.~Chen, G.~Kramer, and H.~Poor, ``Capacity regions and sum-rate
  capacities of vector gaussian interference channels,'' \emph{IEEE
  Transactions on Information Theory}, vol.~56, no.~10, pp. 5030 --5044, oct.
  2010.

\bibitem{bhatia1997}
\BIBentryALTinterwordspacing
R.~Bhatia, \emph{Matrix analysis}, ser. Graduate texts in mathematics.\hskip
  1em plus 0.5em minus 0.4em\relax Springer, 1997. [Online]. Available:
  \url{http://books.google.com/books?id=f0ioPwAACAAJ}
\BIBentrySTDinterwordspacing

\bibitem{freund04ocw}
\BIBentryALTinterwordspacing
R.~Freund, ``15.084j nonlinear programming, spring 2004. (massachusetts
  institute of technology: Mit opencourseware), http://ocw.mit.edu (accessed 16
  jul, 2012). license: Creative commons by-nc-sa.'' [Online]. Available:
  \url{http://ocw.mit.edu/courses/sloan-school-of-management/15-084j-nonlinear-programming-spring-2004/index.htm}
\BIBentrySTDinterwordspacing

\bibitem{aubin2007mathematical}
\BIBentryALTinterwordspacing
J.~Aubin, \emph{Mathematical methods of game and economic theory}, ser. Dover
  books on mathematics.\hskip 1em plus 0.5em minus 0.4em\relax Dover
  Publications, 2007. [Online]. Available:
  \url{http://books.google.com/books?id=COhFPgAACAAJ}
\BIBentrySTDinterwordspacing

\end{thebibliography}

\appendices{}

\section{\label{sec:Proof Lemma marginal}Proof of Lemma \ref{lem:marginal_1}}

Given (\ref{eq:side info}), we have 
\begin{eqnarray*}
\underline{S}_{\left(j-1\right)p}^{n} & = & \left|C_{\left(j-1\right)j}\right|\mathbf{U}_{p\left(j-1\right)}\underline{X}_{\left(j-1\right)}^{n}+\sum_{i=j}^{K}\left|C_{pi}\right|\mathbf{U}_{pi}\underline{X}_{i}^{n}+\underline{W}_{\left(j-1\right)p}^{n}\\
\underline{S}_{\left(j-1\right)\left(p-1\right)}^{n} & = & \left|C_{\left(j-1\right)j}\right|\mathbf{U}_{\left(p-1\right)\left(j-1\right)}\underline{X}_{\left(j-1\right)}^{n}+\sum_{i=j}^{K}\left|C_{\left(p-1\right)i}\right|\mathbf{U}_{\left(p-1\right)i}\underline{X}_{i}^{n}+\underline{W}_{\left(j-1\right)\left(p-1\right)}^{n}\\
 & \vdots\\
\underline{S}_{\left(j-1\right)1}^{n} & = & \left|C_{\left(j-1\right)j}\right|\mathbf{U}_{1\left(j-1\right)}\underline{X}_{\left(j-1\right)}^{n}+\sum_{i=j}^{K}\left|C_{1i}\right|\mathbf{U}_{1i}\underline{X}_{i}^{n}+\underline{W}_{\left(j-1\right)1}^{n}
\end{eqnarray*}
When $\underline{X}_{\left(j-1\right)}^{n}$ is given, it can be subtracted
from $\underline{S}_{\left(j-1\right)p}^{n},\underline{S}_{\left(j-1\right)\left(p-1\right)}^{n},\cdots,\,\underline{S}_{\left(j-1\right)1}^{n}$
to give 
\begin{eqnarray}
\underline{\hat{S}}_{\left(j-1\right)p}^{n} & = & \underline{S}_{\left(j-1\right)p}^{n}-\left|C_{\left(j-1\right)j}\right|\mathbf{U}_{p\left(j-1\right)}\underline{X}_{\left(j-1\right)}^{n}\nonumber \\
 & = & \sum_{i=j}^{K}\left|C_{pi}\right|\mathbf{U}_{pi}\underline{X}_{i}^{n}+\underline{W}_{\left(j-1\right)p}^{n}\label{eq:shat1-1}\\
\underline{\hat{S}}_{\left(j-1\right)\left(p-1\right)}^{n} & = & \underline{S}_{\left(j-1\right)\left(p-1\right)}^{n}-\left|C_{\left(j-1\right)j}\right|\mathbf{U}_{\left(p-1\right)\left(j-1\right)}\underline{X}_{\left(j-1\right)}^{n}\nonumber \\
 & = & \sum_{i=j}^{K}\left|C_{\left(p-1\right)i}\right|\mathbf{U}_{\left(p-1\right)i}\underline{X}_{i}^{n}+\underline{W}_{\left(j-1\right)\left(p-1\right)}^{n}\label{eq:shat2-1}\\
 & \vdots\nonumber \\
\underline{\hat{S}}_{\left(j-1\right)1}^{n} & = & \underline{S}_{\left(j-1\right)1}^{n}-\left|C_{\left(j-1\right)j}\right|\mathbf{U}_{1\left(j-1\right)}\underline{X}_{\left(j-1\right)}^{n}\nonumber \\
 & = & \sum_{i=j}^{K}\left|C_{1i}\right|\mathbf{U}_{1i}\underline{X}_{i}^{n}+\underline{W}_{\left(j-1\right)1}^{n}\label{eq:shat3-1}
\end{eqnarray}
while 
\begin{eqnarray}
\underline{S}_{jp}^{n} & = & \sum_{i=j}^{K}\left|C_{pi}\right|\mathbf{U}_{pi}\underline{X}_{i}^{n}+\underline{W}_{jp}^{n}\label{eq:s1-1}\\
\underline{S}_{j\left(p-1\right)}^{n} & = & \sum_{i=j}^{K}\left|C_{\left(p-1\right)i}\right|\mathbf{U}_{\left(p-1\right)i}\underline{X}_{i}^{n}+\underline{W}_{j\left(p-1\right)}^{n}\label{eq:s2-1}\\
 & \vdots\nonumber \\
\underline{S}_{j1}^{n} & = & \sum_{i=j}^{K}\left|C_{1i}\right|\mathbf{U}_{1i}\underline{X}_{i}^{n}+\underline{W}_{j1}^{n}.\label{eq:s3-1}
\end{eqnarray}

We know that $\left(\underline{Z}_{j},\,\underline{W}_{j\left(j-1\right)},\cdots,\,\underline{W}_{j1}\right)$
are jointly Gaussian random variables, with zero mean and covariance
matrix $\mathbf{K}_{S_{j}}$ equal to: 
\begin{eqnarray*}
 &  & \left(\begin{array}{ccccc}
\mathbf{I} & \mathbf{A}_{j\left(j-1\right)} & \cdots & \mathbf{A}_{j2} & \mathbf{A}_{j1}\\
\mathbf{A}_{j\left(j-1\right)}^{T} & \mathbf{I} & \mathbf{A}_{\left(j-1\right)\left(j-2\right)} & \cdots & \mathbf{A}_{\left(j-1\right)1}\\
\vdots &  & \ddots & \ddots & \vdots\\
\mathbf{A}_{j1}^{T} &  &  & \mathbf{I} & \mathbf{A}_{21}\\
\mathbf{A}_{j1}^{T} & \mathbf{A}_{\left(j-1\right)1}^{T} & \cdots & \mathbf{A}_{21}^{T} & \mathbf{I}
\end{array}\right)\\
\end{eqnarray*}
which is defined in (\ref{eq:side-info cov}). It is clear that the
covariance matrices of the jointly Gaussian random variables $\left(\underline{W}_{\left(j-1\right)p},\,\underline{W}_{\left(j-1\right)\left(p-1\right)},\cdots,\,\underline{W}_{\left(j-1\right)1}\right)$
and $\left(\underline{W}_{jp},\,\underline{W}_{j\left(p-1\right)},\cdots,\,\underline{W}_{j1}\right)$
are the same: 
\begin{eqnarray}
 &  & \cov\left(\underline{W}_{\left(j-1\right)p},\,\underline{W}_{\left(j-1\right)\left(p-1\right)},\cdots,\,\underline{W}_{\left(j-1\right)1}\right)\nonumber \\
 & = & \cov\left(\underline{W}_{jp},\,\underline{W}_{j\left(p-1\right)},\cdots,\,\underline{W}_{j1}\right)\nonumber \\
 & = & \left(\begin{array}{ccccc}
\mathbf{I} & \mathbf{A}_{p\left(p-1\right)} & \cdots & \mathbf{A}_{p2} & \mathbf{A}_{p1}\\
\mathbf{A}_{p\left(p-1\right)}^{T} & \mathbf{I} & \mathbf{A}_{\left(p-1\right)\left(p-2\right)} & \cdots & \mathbf{A}_{\left(p-1\right)1}\\
\vdots &  & \ddots & \ddots & \vdots\\
\mathbf{A}_{p2}^{T} &  &  & \mathbf{I} & \mathbf{A}_{21}\\
\mathbf{A}_{p1}^{T} & \mathbf{A}_{\left(p-1\right)1}^{T} & \cdots & \mathbf{A}_{21}^{T} & \mathbf{I}
\end{array}\right)\label{eq:same cov-1}
\end{eqnarray}
Comparing (\ref{eq:shat1-1})\textasciitilde{}(\ref{eq:shat3-1})
and (\ref{eq:s1-1})\textasciitilde{}(\ref{eq:s3-1}), we can see
that distribution of $\left.\underline{S}_{\left(j-1\right)p}^{n}\right|\underline{S}_{\left(j-1\right)\left(p-1\right)}^{n},\cdots,\,\underline{S}_{\left(j-1\right)1}^{n},\underline{X}_{\left(j-1\right)}^{n}$
and $\left.\underline{S}_{jp}^{n}\right|\underline{S}_{j\left(p-1\right)}^{n},\cdots,\,\underline{S}_{j1}^{n}$
are equal as long as $\left.\underline{W}_{\left(j-1\right)p}^{n}\right|\underline{W}_{\left(j-1\right)\left(p-1\right)}^{n},\cdots,\,\underline{W}_{\left(j-1\right)1}^{n}$
and $\left.\underline{W}_{jp}^{n}\right|\underline{W}_{j\left(p-1\right)}^{n},\cdots,\,\underline{W}_{j1}^{n}$
have the same distribution. Recall that $\underline{W}_{ji}$ is i.i.d.
Gaussian random variables which is independent from the input signals
$\underline{X}^{n}$. Therefore given (\ref{eq:same cov-1}), Lemma
\ref{lem:marginal_1} is proved.

\section{{\normalsize \label{sec:Proof lemma side-info}}Proof of Lemma {\normalsize \ref{lem:side-info}}}

Lemma \ref{lem:side-info} is proved using the following lemma from
\cite{ShangChenKramerIT10}.
\begin{lem}[{\cite[Lemma 4, p5037]{ShangChenKramerIT10}}]
\label{lem:joint gaussian markov}Let $\underline{X}$, $\underline{Y}$
and $\underline{Z}$ be jointly Gaussian vectors. If $\cov\left(\underline{Y}\right)$
is invertible, then $\underline{X}\rightarrow\underline{Y}\rightarrow\underline{Z}$
forms a Markov chain if and only if 
\begin{eqnarray*}
\cov\left(\underline{X},\,\underline{Z}\right) & = & \cov\left(\underline{X},\,\underline{Y}\right)\cov\left(\underline{Y}\right)^{-1}\cov\left(\underline{Y},\,\underline{Z}\right)
\end{eqnarray*}

\end{lem}
Given Lemma \ref{lem:joint gaussian markov} and the fact that $\cov\left(\underline{\hat{Y}}_{jG}\right)$
is invertible, $\underline{X}_{jG}\rightarrow\underline{\hat{Y}}_{jG}\rightarrow\mathbf{\underline{S}}_{j}$
forms a Markov chain if and only if 
\begin{eqnarray}
\lefteqn{} &  & \cov\left(\underline{X}_{jG},\,\mathbf{\underline{S}}_{j}\right)\nonumber \\
 & = & \cov\left(\underline{X}_{jG},\,\underline{\hat{Y}}_{jG}\right)\cov\left(\underline{\hat{Y}}_{jG}\right)^{-1}\cov\left(\underline{\hat{Y}}_{jG},\,\mathbf{\underline{S}}_{j}\right)\label{eq:markov condition}
\end{eqnarray}
Given (\ref{eq:k-user z-channel}), (\ref{eq:side info}) and the
independence of $\underline{W}_{jp}$ and $\underline{X}_{i}$, the
left hand side of (\ref{eq:markov condition}) is
\begin{eqnarray*}
LHS & = & \left(\begin{array}{c}
\left|C_{1j}\right|^{2}\mathbf{V}_{j}\mathbf{U}\left(-\phi_{1j}\right)\\
\left|C_{2j}\right|^{2}\mathbf{V}_{j}\mathbf{U}\left(-\phi_{2j}\right)\\
\vdots\\
\left|C_{\left(j-1\right)j}\right|^{2}\mathbf{V}_{j}\mathbf{U}\left(-\phi_{\left(j-1\right)j}\right)
\end{array}\right)^{T}
\end{eqnarray*}
and the right hand side is 
\begin{eqnarray*}
RHS & = & \left|C_{jj}\right|^{2}\mathbf{V}_{j}\mathbf{U}\left(-\phi_{jj}\right)\left(\sum_{i=j}^{K}\left|C_{ji}\right|^{2}\mathbf{U}\left(\phi_{ji}\right)\mathbf{V}_{i}\mathbf{U}\left(-\phi_{ji}\right)+\mathbf{I}\right)^{-1}\\
 &  & \left(\begin{array}{c}
\sum_{i=j}^{K}\left|C_{ji}\right|^{2}\left|C_{1i}\right|^{2}\mathbf{U}\left(\phi_{ji}\right)\mathbf{V}_{i}\mathbf{U}\left(-\phi_{1i}\right)+\mathbf{A}_{j1}\\
\sum_{i=j}^{K}\left|C_{ji}\right|^{2}\left|C_{2i}\right|^{2}\mathbf{U}\left(\phi_{ji}\right)\mathbf{V}_{i}\mathbf{U}\left(-\phi_{2i}\right)+\mathbf{A}_{j2}\\
\vdots\\
\sum_{i=j+1}^{K}\left|C_{ji}\right|^{2}\left|C_{\left(j-1\right)i}\right|^{2}\mathbf{U}\left(\phi_{ji}\right)\mathbf{V}_{i}\mathbf{U}\left(-\phi_{\left(j-1\right)i}\right)+\mathbf{A}_{j\left(j-1\right)}
\end{array}\right)^{T}
\end{eqnarray*}
In order for $LHS=RHS$, we must have
\begin{eqnarray*}
\lefteqn{} &  & \left|C_{pj}\right|^{2}\mathbf{V}_{j}\mathbf{U}\left(-\phi_{pj}\right)\\
 & = & \left|C_{jj}\right|^{2}\mathbf{V}_{j}\mathbf{U}\left(-\phi_{jj}\right)\left(\sum_{i=j}^{K}\left|C_{ji}\right|^{2}\mathbf{U}\left(\phi_{ji}\right)\mathbf{V}_{i}\mathbf{U}\left(-\phi_{ji}\right)+\mathbf{I}\right)^{-1}\\
 &  & \left(\sum_{i=j}^{K}\left|C_{ji}\right|^{2}\left|C_{pi}\right|^{2}\mathbf{U}\left(\phi_{ji}\right)\mathbf{V}_{i}\mathbf{U}\left(-\phi_{pi}\right)+\mathbf{A}_{jp}\right)
\end{eqnarray*}
Solving the equation above, we have
\begin{eqnarray*}
\mathbf{A}_{jp} & = & \frac{\left|C_{pj}\right|^{2}}{\left|C_{jj}\right|^{2}}\mathbf{U}\left(\phi_{jj}-\phi_{pj}\right)\\
 &  & +\frac{\left|C_{pj}\right|^{2}}{\left|C_{jj}\right|^{2}}\sum_{i=j+1}^{K}\left|C_{ji}\right|^{2}\mathbf{U}\left(\phi_{ji}\right)\mathbf{V}_{i}\mathbf{U}\left(\phi_{jj}-\phi_{pj}-\phi_{ji}\right)\\
 &  & -\sum_{i=j+1}^{K}\left|C_{ji}\right|^{2}\left|C_{pi}\right|^{2}\mathbf{U}\left(\phi_{ji}\right)\mathbf{V}_{i}\mathbf{U}\left(-\phi_{pi}\right)
\end{eqnarray*}

\section{{\normalsize \label{sec:Proof-of-Lemma sufficient}}Proof of Lemma
{\normalsize \ref{lem:sufficient} }}

First, consider the simple case where $\frac{\left|C_{pj}\right|^{2}}{\left|C_{jj}\right|^{2}}=\alpha$,
$\phi_{ji}=0$ and $P_{j}=0$, that is, $\mathbf{K}_{x_{j}}=\mathbf{0}$.
For this case, given (\ref{eq:side-info cov}) we have $\mathbf{A}_{ji}=\mathbf{B}=\left(\begin{array}{cc}
\alpha & 0\\
0 & \alpha
\end{array}\right)$ for all $i,\, j$. It is easy to check that the eigenvalues of $\mathbf{K}_{S_{j}}=\left(\begin{array}{cccc}
\mathbf{I} & \mathbf{B} & \cdots & \mathbf{B}\\
\mathbf{B}^{T} & \mathbf{I} & \cdots & \mathbf{B}\\
\vdots &  & \ddots & \vdots\\
\mathbf{B}^{T} &  & \cdots & \mathbf{I}
\end{array}\right)$ are $\lambda_{1}=1-\alpha$ and $\lambda_{2}=1+\left(j-1\right)\alpha$,
with multiplicity $2\left(j-1\right)$ and 2 respectively. Therefore,
$\mathbf{K}_{S_{j}}$ is positive definite if $0<\alpha<1$. 

Now let us consider the case where $\phi_{ji}$ and $P_{j}$ are small
but non-zero, and $\frac{\left|C_{pj}\right|^{2}}{\left|C_{jj}\right|^{2}}$
are not necessarily equal to $\alpha$. Denote the $\left(p,q\right)th$
element of $\mathbf{B}$ by $b_{pq}$. It is well known that the eigenvalues
of symmetric matrix are locally (Lipschitz) continuous\cite{bhatia1997}
with respect to its elements. Therefore, corresponding to every $\alpha\in\left(0,\,1\right)$,
for any $\hat{\epsilon}>0$, there exist some strictly positive real
numbers $\epsilon_{\alpha}$, $\epsilon_{\alpha}^{\prime}$ and $\epsilon_{\alpha}^{\prime\prime}(\underline{C})$
such that if $\left|\frac{\left|C_{pj}\right|^{2}}{\left|C_{jj}\right|^{2}}-\alpha\right|<\epsilon_{\alpha}$,
$\left|\phi_{ji}\right|<\epsilon_{\alpha}'$, and $P_{j}<\epsilon_{\alpha}''(\underline{C})$
then every eigenvalues $\lambda_{s}$ of $\mathbf{K}_{S_{j}}$ satisfies
$\left|\lambda_{s}-\lambda_{1}\right|<\hat{\epsilon}$ or $\left|\lambda_{s}-\lambda_{2}\right|<\hat{\epsilon}$.
The bound on $P_{j}$ may depend $\underline{C}$ to ensure that the
two last terms in (\ref{eq:side-info cov}) are of bounded variation.
For any $0<\alpha<1$ we can always find some $\hat{\epsilon}>0$
that guarantees $\lambda_{s}>0$, and $\mathbf{K}_{S_{j}}$ is positive
definite as a result.

\section{{\normalsize \label{sec:z-channel sum rate}}Proof of Theorem {\normalsize \ref{thm:K-user sum rate}}}

First we state a useful result from \cite{ShangChenKramerIT10}.
\begin{lem}
\label{lem:gaussian x}(\cite[Lemma 2]{ShangChenKramerIT10}) Let
$\underline{X}^{n}=\left(\underline{X}_{1},\cdots,\underline{X}_{n}\right)$
and $\underline{Y}^{n}=\left(\underline{Y}_{1},\cdots,\underline{Y}_{n}\right)$
be two sequences of random vectors, and let $\underline{X}{}_{G}^{\prime}$,
$\underline{X}{}_{G}$, $\underline{Y}{}_{G}^{\prime}$, and $\underline{Y}{}_{G}$
be Gaussian vectors with covariance matrices satisfying 
\[
\cov\left(\begin{array}{c}
\underline{X}{}_{G}^{\prime}\\
\underline{Y}{}_{G}^{\prime}
\end{array}\right)=\frac{1}{n}\sum_{i=1}^{n}\cov\left(\begin{array}{c}
\underline{X}_{i}\\
\underline{Y}_{i}
\end{array}\right)\preceq\cov\left(\begin{array}{c}
\underline{X}{}_{G}\\
\underline{Y}{}_{G}
\end{array}\right)
\]
then we have 
\begin{eqnarray*}
h\left(\underline{X}^{n}\right) & \leq nh\left(\underline{X}{}_{G}^{\prime}\right) & \leq nh\left(\underline{X}{}_{G}\right)\\
h\left(\left.\underline{Y}^{n}\right|\underline{X}^{n}\right) & \leq nh\left(\left.\underline{Y}{}_{G}^{\prime}\right|\underline{X}{}_{G}^{\prime}\right) & \leq nh\left(\left.\underline{Y}{}_{G}\right|\underline{X}{}_{G}\right)
\end{eqnarray*}

\end{lem}
By Fano's inequality, the sum capacity of the generalized Z-channel
(\ref{eq:k-user z-channel}) must satisfy 
\begin{eqnarray*}
\lefteqn{n\sum_{j=1}^{K}R_{j}-n\epsilon}\\
 & \stackrel{(a)}{\leq} & I\left(\underline{X}_{1}^{n};\,\underline{\hat{Y}}_{1}^{n}\right)+\sum_{j=2}^{K}I\left(\underline{X}_{j}^{n};\,\underline{\hat{Y}}_{j}^{n},\,\mathbf{\underline{S}}_{j}\right)\\
 & \stackrel{(b)}{=} & h\left(\underline{\hat{Y}}_{1}^{n}\right)-h\left(\left.\underline{\hat{Y}}_{1}^{n}\right|\underline{X}_{1}^{n}\right)\\
 &  & +\sum_{j=2}^{K}I\left(\underline{X}_{j}^{n};\,\mathbf{\underline{S}}_{j}\right)+\sum_{j=2}^{K}I\left(\left.\underline{X}_{j}^{n};\,\underline{\hat{Y}}_{j}^{n}\right|\mathbf{\underline{S}}_{j}\right)\\
 & \stackrel{(c)}{=} & h\left(\underline{\hat{Y}}_{1}^{n}\right)-h\left(\left.\underline{\hat{Y}}_{1}^{n}\right|\underline{X}_{1}^{n}\right)\\
 &  & +\sum_{j=2}^{K}\sum_{p=1}^{j-1}I\left(\left.\underline{X}_{j}^{n};\,\underline{S}_{jp}\right|\underline{S}_{j\left(p-1\right)}^{n},\cdots,\,\underline{S}_{j1}^{n}\right)\\
 &  & +\sum_{j=2}^{K}I\left(\left.\underline{X}_{j}^{n};\,\underline{\hat{Y}}_{j}^{n}\right|\mathbf{\underline{S}}_{j}\right)\\
 & \stackrel{(d)}{=} & h\left(\underline{\hat{Y}}_{1}^{n}\right)-h\left(\left.\underline{\hat{Y}}_{1}^{n}\right|\underline{X}_{1}^{n}\right)\\
 &  & +\sum_{j=2}^{K}\sum_{p=1}^{j-1}\left(h\left(\left.\underline{S}_{jp}^{n}\right|\underline{S}_{j\left(p-1\right)}^{n},\cdots,\,\underline{S}_{j1}^{n}\right)\right.\\
 &  & \left.-h\left(\left.\underline{S}_{jp}^{n}\right|\underline{S}_{j\left(p-1\right)}^{n},\cdots,\,\underline{S}_{j1}^{n},\underline{X}_{j}^{n}\right)\right)\\
 &  & +\sum_{j=2}^{K}\left(h\left(\left.\underline{\hat{Y}}_{j}^{n}\right|\underline{S}_{j\left(j-1\right)}^{n},\underline{S}_{j\left(j-2\right)}^{n},\cdots,\,\underline{S}_{j1}^{n}\right)\right.\\
 &  & \left.-h\left(\left.\underline{\hat{Y}}_{j}^{n}\right|\underline{S}_{j\left(j-1\right)}^{n},\underline{S}_{j\left(j-2\right)}^{n},\cdots,\,\underline{S}_{j1}^{n},\underline{X}_{j}^{n}\right)\right)\\
 & \stackrel{(e)}{\leq} & nh\left(\underline{\hat{Y}}_{1G}\right)-h\left(\left.\underline{\hat{Y}}_{1}^{n}\right|\underline{X}_{1}^{n}\right)\\
 &  & +h\left(\underline{S}_{21}^{n}\right)-h\left(\left.\underline{S}_{21}^{n}\right|\underline{X}_{2}^{n}\right)\\
 &  & +\sum_{j=3}^{K}\sum_{p=j-1}^{j-1}h\left(\left.\underline{S}_{jp}^{n}\right|\underline{S}_{j\left(p-1\right)}^{n},\cdots,\,\underline{S}_{j1}^{n}\right)\\
 &  & +\sum_{j=3}^{K}\sum_{p=1}^{j-2}h\left(\left.\underline{S}_{jp}^{n}\right|\underline{S}_{j\left(p-1\right)}^{n},\cdots,\,\underline{S}_{j1}^{n}\right)\\
 &  & -\sum_{j=3}^{K-1}\sum_{p=1}^{j-1}h\left(\left.\underline{S}_{jp}^{n}\right|\underline{S}_{j\left(p-1\right)}^{n},\cdots,\,\underline{S}_{j1}^{n},\underline{X}_{j}^{n}\right)\\
 &  & -\sum_{j=K}^{K}\sum_{p=1}^{j-1}h\left(\left.\underline{S}_{jp}^{n}\right|\underline{S}_{j\left(p-1\right)}^{n},\cdots,\,\underline{S}_{j1}^{n},\underline{X}_{j}^{n}\right)\\
 &  & +\sum_{j=2}^{K}h\left(\left.\underline{\hat{Y}}_{j}^{n}\right|\underline{S}_{j\left(j-1\right)}^{n},\underline{S}_{j\left(j-2\right)}^{n},\cdots,\,\underline{S}_{j1}^{n}\right)\\
 &  & -\sum_{j=2}^{K}h\left(\left.\underline{\hat{Y}}_{j}^{n}\right|\underline{S}_{j\left(j-1\right)}^{n},\underline{S}_{j\left(j-2\right)}^{n},\cdots,\,\underline{S}_{j1}^{n},\underline{X}_{j}^{n}\right)
\end{eqnarray*}
\begin{eqnarray*}
 & \stackrel{(f)}{=} & nh\left(\underline{\hat{Y}}_{1G}\right)-h\left(\left.\underline{\hat{Y}}_{1}^{n}\right|\underline{X}_{1}^{n}\right)\\
 &  & +h\left(\underline{S}_{21}^{n}\right)+\sum_{j=3}^{K}h\left(\left.\underline{S}_{j\left(j-1\right)}^{n}\right|\underline{S}_{j\left(j-2\right)}^{n},\cdots,\,\underline{S}_{j1}^{n}\right)\\
 &  & +\sum_{j=3}^{K}\sum_{p=1}^{j-2}\left(h\left(\left.\underline{S}_{jp}^{n}\right|\underline{S}_{j\left(p-1\right)}^{n},\cdots,\,\underline{S}_{j1}^{n}\right)\right.\\
 &  & \left.-h\left(\left.\underline{S}_{\left(j-1\right)p}^{n}\right|\underline{S}_{\left(j-1\right)\left(p-1\right)}^{n},\cdots,\,\underline{S}_{\left(j-1\right)1}^{n},\underline{X}_{\left(j-1\right)}^{n}\right)\right)\\
 &  & -\sum_{p=1}^{K-1}h\left(\left.\underline{S}_{Kp}^{n}\right|\underline{S}_{K\left(p-1\right)}^{n},\cdots,\,\underline{S}_{K1}^{n},\underline{X}_{K}^{n}\right)\\
 &  & +\sum_{j=2}^{K}h\left(\left.\underline{\hat{Y}}_{j}^{n}\right|\underline{S}_{j\left(j-1\right)}^{n},\underline{S}_{j\left(j-2\right)}^{n},\cdots,\,\underline{S}_{j1}^{n}\right)\\
 &  & -\sum_{j=2}^{K}h\left(\left.\underline{\hat{Y}}_{j}^{n}\right|\underline{S}_{j\left(j-1\right)}^{n},\underline{S}_{j\left(j-2\right)}^{n},\cdots,\,\underline{S}_{j1}^{n},\underline{X}_{j}^{n}\right)\\
 & \stackrel{(g)}{=} & nh\left(\underline{\hat{Y}}_{1G}\right)-h\left(\left.\underline{\hat{Y}}_{1}^{n}\right|\underline{X}_{1}^{n}\right)\\
 &  & +h\left(\underline{S}_{21}^{n}\right)+\sum_{j=3}^{K}h\left(\left.\underline{S}_{j\left(j-1\right)}^{n}\right|\underline{S}_{j\left(j-2\right)}^{n},\cdots,\,\underline{S}_{j1}^{n}\right)\\
 &  & -\sum_{p=1}^{K-1}h\left(\left.\underline{S}_{Kp}^{n}\right|\underline{S}_{K\left(p-1\right)}^{n},\cdots,\,\underline{S}_{K1}^{n},\underline{X}_{K}^{n}\right)\\
 &  & +\sum_{j=2}^{K}h\left(\left.\underline{\hat{Y}}_{j}^{n}\right|\underline{S}_{j\left(j-1\right)}^{n},\underline{S}_{j\left(j-2\right)}^{n},\cdots,\,\underline{S}_{j1}^{n}\right)\\
 &  & -\sum_{j=2}^{K}h\left(\left.\underline{\hat{Y}}_{j}^{n}\right|\underline{S}_{j\left(j-1\right)}^{n},\underline{S}_{j\left(j-2\right)}^{n},\cdots,\,\underline{S}_{j1}^{n},\underline{X}_{j}^{n}\right)\\
 & \stackrel{(h)}{=} & nh\left(\underline{\hat{Y}}_{1G}\right)+\sum_{j=3}^{K}h\left(\left.\underline{S}_{j\left(j-1\right)}^{n}\right|\underline{S}_{j\left(j-2\right)}^{n},\cdots,\,\underline{S}_{j1}^{n}\right)\\
 &  & -nh\left(\underline{W}_{K\left(K-1\right)},\cdots,\,\underline{W}_{K1}\right)\\
 &  & +\sum_{j=2}^{K}h\left(\left.\underline{\hat{Y}}_{j}^{n}\right|\underline{S}_{j\left(j-1\right)}^{n},\underline{S}_{j\left(j-2\right)}^{n},\cdots,\,\underline{S}_{j1}^{n}\right)\\
 &  & -\sum_{j=2}^{K}h\left(\left.\underline{\hat{Y}}_{j}^{n}\right|\underline{S}_{j\left(j-1\right)}^{n},\underline{S}_{j\left(j-2\right)}^{n},\cdots,\,\underline{S}_{j1}^{n},\underline{X}_{j}^{n}\right)
\end{eqnarray*}

\begin{eqnarray*}
 & \stackrel{(i)}{=} & nh\left(\underline{\hat{Y}}_{1G}\right)-nh\left(\underline{W}_{K\left(K-1\right)},\cdots,\,\underline{W}_{K1}\right)\\
 &  & +\sum_{j=2}^{K}h\left(\left.\underline{\hat{Y}}_{j}^{n}\right|\underline{S}_{j\left(j-1\right)}^{n},\underline{S}_{j\left(j-2\right)}^{n},\cdots,\,\underline{S}_{j1}^{n}\right)\\
 &  & -nh\left(\left.\underline{N}_{K}\right|\underline{W}_{K\left(K-1\right)},\cdots,\,\underline{W}_{K1}\right)\\
 & \stackrel{(j)}{=} & nh\left(\underline{\hat{Y}}_{1G}\right)-nh\left(\underline{N}_{K},\,\underline{W}_{K\left(K-1\right)},\cdots,\,\underline{W}_{K1}\right)\\
 &  & +\sum_{j=2}^{K}h\left(\left.\underline{\hat{Y}}_{j}^{n}\right|\underline{S}_{j\left(j-1\right)}^{n},\underline{S}_{j\left(j-2\right)}^{n},\cdots,\,\underline{S}_{j1}^{n}\right)\\
 & \stackrel{(k)}{\leq} & nh\left(\underline{\hat{Y}}_{1G}\right)-nh\left(\underline{N}_{K},\,\underline{W}_{K\left(K-1\right)},\cdots,\,\underline{W}_{K1}\right)\\
 &  & +n\sum_{j=2}^{K}h\left(\left.\underline{\hat{Y}}_{jG}\right|\underline{S}_{j\left(j-1\right)G},\underline{S}_{j\left(j-2\right)G},\cdots,\,\underline{S}_{j1G}\right)\\
 & \stackrel{(l)}{=} & nh\left(\underline{\hat{Y}}_{1G}\right)-nh\left(\underline{N}_{K},\,\underline{W}_{K\left(K-1\right)},\cdots,\,\underline{W}_{K1}\right)\\
 &  & +n\sum_{j=2}^{K}h\left(\underline{\hat{Y}}_{jG}\right)\\
 &  & -n\sum_{j=2}^{K}h\left(\underline{S}_{j\left(j-1\right)G},\underline{S}_{j\left(j-2\right)G},\cdots,\,\underline{S}_{j1G}\right)\\
 &  & +n\sum_{j=2}^{K}h\left(\left.\underline{S}_{j\left(j-1\right)G},\underline{S}_{j\left(j-2\right)G},\cdots,\,\underline{S}_{j1G}\right|\underline{\hat{Y}}_{jG}\right)
\end{eqnarray*}

\begin{eqnarray*}
 & \stackrel{(m)}{=} & nh\left(\underline{\hat{Y}}_{1G}\right)-nh\left(\underline{N}_{K},\,\underline{W}_{K\left(K-1\right)},\cdots,\,\underline{W}_{K1}\right)\\
 &  & +n\sum_{j=2}^{K}h\left(\underline{\hat{Y}}_{jG}\right)\\
 &  & -n\sum_{j=2}^{K}h\left(\underline{S}_{j\left(j-1\right)G},\underline{S}_{j\left(j-2\right)G},\cdots,\,\underline{S}_{j1G}\right)\\
 &  & +n\sum_{j=2}^{K-1}h\left(\left.\underline{S}_{j\left(j-1\right)G},\underline{S}_{j\left(j-2\right)G},\cdots,\,\underline{S}_{j1G}\right|\underline{\hat{Y}}_{jG}\right)\\
 &  & +nh\left(\left.\underline{S}_{K\left(K-1\right)G},\underline{S}_{K\left(K-2\right)G},\cdots,\,\underline{S}_{K1G}\right|\underline{\hat{Y}}_{KG}\right)\\
 & \stackrel{(n)}{=} & nh\left(\underline{\hat{Y}}_{1G}\right)-nh\left(\underline{N}_{K},\,\underline{W}_{K\left(K-1\right)},\cdots,\,\underline{W}_{K1}\right)\\
 &  & +n\sum_{j=2}^{K}h\left(\underline{\hat{Y}}_{jG}\right)\\
 &  & -n\sum_{j=2}^{K}h\left(\underline{S}_{j\left(j-1\right)G},\underline{S}_{j\left(j-2\right)G},\cdots,\,\underline{S}_{j1G}\right)\\
 &  & +n\sum_{j=2}^{K-1}h\left(\left.\underline{S}_{\left(j+1\right)\left(j-1\right)G},\underline{S}_{\left(j+1\right)\left(j-2\right)G},\cdots,\,\underline{S}_{\left(j+1\right)1G}\right|\underline{S}_{\left(j+1\right)jG}\right)\\
 &  & +nh\left(\left.\underline{W}_{K\left(K-1\right)},\underline{W}_{K\left(K-2\right)},\cdots,\,\underline{W}_{K1}\right|\underline{N}_{K}\right)\\
 & = & nh\left(\underline{\hat{Y}}_{1G}\right)-nh\left(\underline{N}_{K},\,\underline{W}_{K\left(K-1\right)},\cdots,\,\underline{W}_{K1}\right)\\
 &  & +n\sum_{j=2}^{K}h\left(\underline{\hat{Y}}_{jG}\right)\\
 &  & -n\sum_{j=2}^{K}h\left(\underline{S}_{j\left(j-1\right)G},\underline{S}_{j\left(j-2\right)G},\cdots,\,\underline{S}_{j1G}\right)\\
 &  & +n\sum_{j=3}^{K}h\left(\left.\underline{S}_{j\left(j-2\right)G},\underline{S}_{j\left(j-3\right)G},\cdots,\,\underline{S}_{j1G}\right|\underline{S}_{j\left(j-1\right)G}\right)\\
 &  & +nh\left(\left.\underline{W}_{K\left(K-1\right)},\underline{W}_{K\left(K-2\right)},\cdots,\,\underline{W}_{K1}\right|\underline{N}_{K}\right)\\
 & = & nh\left(\underline{\hat{Y}}_{1G}\right)-nh\left(\underline{N}_{K},\,\underline{W}_{K\left(K-1\right)},\cdots,\,\underline{W}_{K1}\right)\\
 &  & +n\sum_{j=2}^{K}h\left(\underline{\hat{Y}}_{jG}\right)-h\left(\underline{S}_{21G}\right)\\
 &  & -n\sum_{j=3}^{K}\left(h\left(\underline{S}_{j\left(j-1\right)G},\underline{S}_{j\left(j-2\right)G},\cdots,\,\underline{S}_{j1G}\right)\right.\\
 &  & -\left.h\left(\left.\underline{S}_{j\left(j-2\right)G},\underline{S}_{j\left(j-3\right)G},\cdots,\,\underline{S}_{j1G}\right|\underline{S}_{j\left(j-1\right)G}\right)\right)\\
 &  & +nh\left(\left.\underline{W}_{K\left(K-1\right)},\underline{W}_{K\left(K-2\right)},\cdots,\,\underline{W}_{K1}\right|\underline{N}_{K}\right)
\end{eqnarray*}
\begin{eqnarray*}
 & = & nh\left(\underline{\hat{Y}}_{1G}\right)-nh\left(\underline{N}_{K},\,\underline{W}_{K\left(K-1\right)},\cdots,\,\underline{W}_{K1}\right)\\
 &  & +n\sum_{j=2}^{K}h\left(\underline{\hat{Y}}_{jG}\right)-h\left(\underline{S}_{21G}\right)\\
 &  & -n\sum_{j=3}^{K}h\left(\underline{S}_{j\left(j-1\right)G}\right)\\
 &  & +nh\left(\left.\underline{W}_{K\left(K-1\right)},\underline{W}_{K\left(K-2\right)},\cdots,\,\underline{W}_{K1}\right|\underline{N}_{K}\right)\\
 & = & nh\left(\underline{\hat{Y}}_{1G}\right)-nh\left(\underline{N}_{K}\right)\\
 &  & +n\sum_{j=2}^{K}h\left(\underline{\hat{Y}}_{jG}\right)-h\left(\underline{S}_{21G}\right)\\
 &  & -n\sum_{j=3}^{K}h\left(\underline{S}_{j\left(j-1\right)G}\right)\\
 & = & n\sum_{j=1}^{K-1}\left(h\left(\underline{\hat{Y}}_{jG}\right)-h\left(\underline{S}_{\left(j+1\right)jG}\right)\right)\\
 &  & +nh\left(\underline{\hat{Y}}_{KG}\right)-nh\left(\underline{N}_{K}\right)\\
 & = & n\sum_{j=1}^{K}I\left(\underline{X}_{jG};\underline{\hat{Y}}_{jG}\right)
\end{eqnarray*}

\textbf{(a)} is from Fano's inequality.

\textbf{(b)} is from the expansion of mutual information: $I\left(\underline{X}_{1}^{n};\,\underline{\hat{Y}}_{1}^{n}\right)=h\left(\underline{\hat{Y}}_{1}^{n}\right)-h\left(\left.\underline{\hat{Y}}_{1}^{n}\right|\underline{X}_{1}^{n}\right)$,
and the chain rule which gives $I\left(\underline{X}_{j}^{n};\,\underline{\hat{Y}}_{j}^{n},\,\mathbf{\underline{S}}_{j}\right)=I\left(\underline{X}_{j}^{n};\,\mathbf{\underline{S}}_{j}\right)+I\left(\left.\underline{X}_{j}^{n};\,\underline{\hat{Y}}_{j}^{n}\right|\mathbf{\underline{S}}_{j}\right)$.. 

\textbf{(c)} is from the chain rule, which gives $I\left(\underline{X}_{j}^{n};\,\mathbf{\underline{S}}_{j}\right)=\sum_{p=1}^{j-1}I\left(\left.\underline{X}_{j}^{n};\,\underline{S}_{jp}\right|\underline{S}_{j\left(p-1\right)}^{n},\cdots,\,\underline{S}_{j1}^{n}\right)$.

\textbf{(d)} is from the expansion of mutual information. 

\textbf{(e)} is from the inequality $h\left(\underline{\hat{Y}}_{1}^{n}\right)\leq nh\left(\underline{\hat{Y}}_{1G}\right)$.
It holds because Gaussian random variable maximize entropy under given
power constraint, and line 2 to line 6 in (e) is equivalent to line
2 and line 3 in (d). 

\textbf{(f)} is from the following equation: 
\begin{eqnarray}
 &  & -h\left(\left.\underline{S}_{21}^{n}\right|\underline{X}_{2}^{n}\right)-\sum_{j=3}^{K-1}\sum_{p=1}^{j-1}h\left(\left.\underline{S}_{jp}^{n}\right|\underline{S}_{j\left(p-1\right)}^{n},\cdots,\,\underline{S}_{j1}^{n},\underline{X}_{j}^{n}\right)\nonumber \\
 & = & -\sum_{j=3}^{K}\sum_{p=1}^{j-1}h\left(\left.\underline{S}_{\left(j-1\right)p}^{n}\right|\underline{S}_{\left(j-1\right)\left(p-1\right)}^{n},\cdots,\,\underline{S}_{\left(j-1\right)1}^{n},\underline{X}_{\left(j-1\right)}^{n}\right).\label{eq:e1}
\end{eqnarray}

\textbf{(g)} is from Lemma \ref{lem:marginal_1}. Because random variables
$\left.\underline{S}_{\left(j-1\right)p}^{n}\right|\underline{S}_{\left(j-1\right)\left(p-1\right)}^{n},\cdots,\,\underline{S}_{\left(j-1\right)1}^{n},\underline{X}_{\left(j-1\right)}^{n}$
and $\left.\underline{S}_{jp}^{n}\right|\underline{S}_{j\left(p-1\right)}^{n},\cdots,\,\underline{S}_{j1}^{n}$
have the same marginal distribution, $h\left(\left.\underline{S}_{\left(j-1\right)p}^{n}\right|\underline{S}_{\left(j-1\right)\left(p-1\right)}^{n},\cdots,\,\underline{S}_{\left(j-1\right)1}^{n},\underline{X}_{\left(j-1\right)}^{n}\right)$
and $h\left(\left.\underline{S}_{jp}^{n}\right|\underline{S}_{j\left(p-1\right)}^{n},\cdots,\,\underline{S}_{j1}^{n}\right)$
are equal, which gives 
\begin{eqnarray*}
 & \sum_{j=3}^{K}\sum_{p=1}^{j-2}\left(h\left(\left.\underline{S}_{jp}^{n}\right|\underline{S}_{j\left(p-1\right)}^{n},\cdots,\,\underline{S}_{j1}^{n}\right)\right.\\
 & \left.-h\left(\left.\underline{S}_{\left(j-1\right)p}^{n}\right|\underline{S}_{\left(j-1\right)\left(p-1\right)}^{n},\cdots,\,\underline{S}_{\left(j-1\right)1}^{n},\underline{X}_{\left(j-1\right)}^{n}\right)\right) & =0.
\end{eqnarray*}

\textbf{(h)} Given $\underline{S}_{Kp}=\left|C_{pK}\right|\mathbf{U}_{pK}\underline{X}_{K}+\underline{W}_{Kp}$
, the summation in the third line after (g) gives 
\begin{eqnarray}
 &  & \sum_{p=1}^{K-1}h\left(\left.\underline{S}_{Kp}^{n}\right|\underline{S}_{K\left(p-1\right)}^{n},\cdots,\,\underline{S}_{K1}^{n},\underline{X}_{K}^{n}\right)\nonumber \\
 & = & \sum_{p=1}^{K-1}h\left(\left.\underline{W}_{Kp}^{n}\right|\underline{W}_{K\left(p-1\right)}^{n},\cdots,\,\underline{W}_{K1}^{n}\right)\\
 & = & h\left(\underline{W}_{K\left(K-1\right)}^{n},\cdots,\,\underline{W}_{K1}^{n}\right)\\
 & = & nh\left(\underline{W}_{K\left(K-1\right)},\cdots,\,\underline{W}_{K1}\right)\label{eq:g1}
\end{eqnarray}
It is also easy to see that $\underline{S}_{21}^{n}$ and $\left.\underline{\hat{Y}}_{1}^{n}\right|\underline{X}_{1}^{n}$
have same marginal distribution, therefore 
\begin{eqnarray}
h\left(\underline{S}_{21}^{n}\right)-h\left(\left.\underline{\hat{Y}}_{1}^{n}\right|\underline{X}_{1}^{n}\right) & = & 0\label{eq:g4}
\end{eqnarray}

\textbf{(i)} Now combine the second and the last terms after (h):
\begin{eqnarray}
 &  & \sum_{j=3}^{K}h\left(\left.\underline{S}_{j\left(j-1\right)}^{n}\right|\underline{S}_{j\left(j-2\right)}^{n},\cdots,\,\underline{S}_{j1}^{n}\right)-\sum_{j=2}^{K}h\left(\left.\underline{\hat{Y}}_{j}^{n}\right|\underline{S}_{j\left(j-1\right)}^{n},\underline{S}_{j\left(j-2\right)}^{n},\cdots,\,\underline{S}_{j1}^{n},\underline{X}_{j}^{n}\right)\nonumber \\
 & = & \sum_{j=3}^{K}h\left(\left.\underline{S}_{j\left(j-1\right)}^{n}\right|\underline{S}_{j\left(j-2\right)}^{n},\cdots,\,\underline{S}_{j1}^{n}\right)-\sum_{j=3}^{K}h\left(\left.\underline{\hat{Y}}_{\left(j-1\right)}^{n}\right|\underline{S}_{\left(j-1\right)\left(j-2\right)}^{n},\cdots,\,\underline{S}_{\left(j-1\right)1}^{n},\underline{X}_{\left(j-1\right)}^{n}\right)\nonumber \\
 &  & -h\left(\left.\underline{\hat{Y}}_{K}^{n}\right|\underline{S}_{K\left(K-1\right)}^{n},\cdots,\,\underline{S}_{K1}^{n},\underline{X}_{K}^{n}\right)\\
 & \stackrel{(h-1)}{=} & -h\left(\left.\underline{\hat{Y}}_{K}^{n}\right|\underline{S}_{K\left(K-1\right)}^{n},\cdots,\,\underline{S}_{K1}^{n},\underline{X}_{K}^{n}\right)\\
 & = & nh\left(\left.\underline{N}_{K}\right|\underline{W}_{K\left(K-1\right)},\cdots,\,\underline{W}_{K1}\right)\label{eq:h1}
\end{eqnarray}
(h-1) is from Lemma \ref{lem:marginal 2}. Given that random variables
$\left.\underline{\hat{Y}}_{j-1}^{n}\right|\underline{S}_{\left(j-1\right)\left(j-2\right)}^{n},\cdots,\,\underline{S}_{\left(j-1\right)1}^{n},\underline{X}_{\left(j-1\right)}^{n}$
and $\left.\underline{S}_{j\left(j-1\right)}^{n}\right|\underline{S}_{j\left(j-2\right)}^{n},\cdots,\,\underline{S}_{j1}^{n}$
have the same marginal distribution, we have 
\[
h\left(\left.\underline{S}_{j\left(j-1\right)}^{n}\right|\underline{S}_{j\left(j-2\right)}^{n},\cdots,\,\underline{S}_{j1}^{n}\right)=h\left(\left.\underline{\hat{Y}}_{\left(j-1\right)}^{n}\right|\underline{S}_{\left(j-1\right)\left(j-2\right)}^{n},\cdots,\,\underline{S}_{\left(j-1\right)1}^{n},\underline{X}_{\left(j-1\right)}^{n}\right),
\]

\textbf{(j)} From chain rule of entropy, we know that 
\begin{eqnarray}
 &  & h\left(\underline{W}_{K\left(K-1\right)},\cdots,\,\underline{W}_{K1}\right)+h\left(\left.\underline{N}_{K}\right|\underline{W}_{K\left(K-1\right)},\cdots,\,\underline{W}_{K1}\right)\nonumber \\
 & = & h\left(\underline{N}_{K},\,\underline{W}_{K\left(K-1\right)},\cdots,\,\underline{W}_{K1}\right).\label{eq:i1}
\end{eqnarray}

\textbf{(k)} is from Lemma \ref{lem:gaussian x}.

\textbf{(l)} is from the formula $h\left(\left.X\right|Y\right)=h\left(X\right)+h\left(\left.Y\right|X\right)-h\left(Y\right)$.

\textbf{(m)} is from 
\begin{eqnarray*}
 &  & \sum_{j=2}^{K}h\left(\left.\underline{S}_{j\left(j-1\right)G},\underline{S}_{j\left(j-2\right)G},\cdots,\,\underline{S}_{j1G}\right|\underline{\hat{Y}}_{jG}\right)\\
 & = & +\sum_{j=2}^{K-1}h\left(\left.\underline{S}_{j\left(j-1\right)G},\underline{S}_{j\left(j-2\right)G},\cdots,\,\underline{S}_{j1G}\right|\underline{\hat{Y}}_{jG}\right)\\
 &  & +h\left(\left.\underline{S}_{K\left(K-1\right)G},\underline{S}_{K\left(K-2\right)G},\cdots,\,\underline{S}_{K1G}\right|\underline{\hat{Y}}_{KG}\right)
\end{eqnarray*}

\textbf{(n) }Combining Lemma \ref{lem:side-info} and Lemma \ref{lem:sufficient},
we know that for channels in $\mathcal{C}_{\alpha},$ if the power
constraint $P_{j}$ satisfies $P_{j}\leq\epsilon_{\alpha}^{\prime\prime}$,
then 
\begin{eqnarray}
 & \underline{X}_{jG}\rightarrow & \underline{\hat{Y}}_{jG}\rightarrow\mathbf{\underline{S}}_{j}\label{eq:markov 1-1}
\end{eqnarray}
form a Markov chain, and the following equality holds:
\begin{eqnarray}
 &  & h\left(\left.\underline{S}_{j\left(j-1\right)G},\underline{S}_{j\left(j-2\right)G},\cdots,\,\underline{S}_{j1G}\right|\underline{\hat{Y}}_{jG}\right)\nonumber \\
 & = & h\left(\left.\underline{S}_{j\left(j-1\right)G},\underline{S}_{j\left(j-2\right)G},\cdots,\,\underline{S}_{j1G}\right|\underline{\hat{Y}}_{jG},\underline{X}_{jG}\right)\nonumber \\
 & = & h\left(\left.\underline{S}_{\left(j+1\right)\left(j-1\right)G},\underline{S}_{\left(j+1\right)\left(j-2\right)G},\cdots,\,\underline{S}_{\left(j+1\right)1G}\right|\underline{S}_{\left(j+1\right)jG}\right)\label{eq:l1}
\end{eqnarray}

We can conclude that the achievable sum capacity of the generalized
Z-channel must satisfy 
\begin{eqnarray}
\sum_{j=1}^{K}R_{j} & \leq & \max_{\begin{array}{c}
\trace\left(\mathbf{V}_{j}\right)\leq\snr_{j}\\
\mathbf{V}_{j}\succeq\mathbf{0},\, j=1,\cdots,K
\end{array}}\sum_{j=1}^{K}I\left(\underline{X}_{jG};\underline{\hat{Y}}_{jG}\right)\label{eq:K-user sum rate bound-1}\\
 & = & \max_{\begin{array}{c}
\trace\left(\mathbf{V}_{j}\right)\leq\snr_{j}\\
\mathbf{V}_{j}\succeq\mathbf{0},\, j=1,\cdots,K
\end{array}}\sum_{j=1}^{K}\log\left|\left(\mathbf{I}+\sum_{i=j}^{K}\left|C_{ji}\right|^{2}\mathbf{V}_{i}\right)\left(\mathbf{I}+\sum_{i=j+1}^{K}\left|C_{ji}\right|^{2}\mathbf{V}_{i}\right)^{-1}\right|
\end{eqnarray}

Notice that for Z-channel, this sum capacity outer bound is achievable
because the expression above is identical to the sum capacity achieved
by treating interference as noise. Since the generalized Z-channel
is obtained by eliminating some of the interference links from the
interference channel, (\ref{eq:K-user sum rate bound-1}) is an outer
bound for the sum capacity of the interference channel. Theorem \ref{thm:K-user sum rate}
is proved.

\section{\label{sec:Proof-of-sum slope outer}Proof of Theorem \ref{thm:sum slope}}

In Theorem \ref{thm:K-user sum rate}, we have proved that the sum
capacity (\ref{eq:K-user sum rate bound}) of the generalized Z-channel
is achieved by i.i.d. Gaussian input, 
\begin{eqnarray}
\sum_{j=1}^{K}R_{j} & \leq & \max_{\begin{array}{c}
\trace\left(\mathbf{V}_{j}\right)\leq P_{j}\\
\mathbf{V}_{j}\succeq\mathbf{0},\, j=1,\cdots,K
\end{array}}\sum_{j=1}^{K}I\left(\underline{X}_{jG};\underline{\hat{Y}}_{jG}\right)\label{eq:K-user sum rate bound-1-1}\\
 & = & \max_{\begin{array}{c}
\trace\left(\mathbf{V}_{j}\right)\leq P_{j}\\
\mathbf{V}_{j}\succeq\mathbf{0},\, j=1,\cdots,K
\end{array}}\sum_{j=1}^{K}\log\left|\left(\mathbf{I}+\sum_{i=j}^{K}\left|C_{ji}\right|^{2}\mathbf{V}_{i}\right)\left(\mathbf{I}+\sum_{i=j+1}^{K}\left|C_{ji}\right|^{2}\mathbf{V}_{i}\right)^{-1}\right|
\end{eqnarray}
Define the normalized covariance matrix $\hat{\mathbf{V}}_{j}=\frac{\mathbf{V}_{j}}{P_{j}}$,
$\trace\left(\hat{\mathbf{V}}_{j}\right)=1$. Consider the equal power
constraint where $P_{j}=\nicefrac{P_{sum}}{K}$ for all users. 

For an expression of the form $\log\left|\mathbf{I}+x\mathbf{A}\right|$,
let the eigenvalue of matrix $\mathbf{A}$ be $0\leq\lambda_{i}\left(\mathbf{A}\right)<\infty$.
Then 
\begin{eqnarray}
\log\left|\mathbf{I}+x\mathbf{A}\right| & = & \sum_{i=1}^{n}\log\left(1+x\lambda_{i}\left(\mathbf{A}\right)\right)\nonumber \\
 & = & \sum_{i=1}^{n}\left(x\lambda_{i}\left(\mathbf{A}\right)-\frac{1}{2}x^{2}\lambda_{i}^{2}\left(\mathbf{A}\right)+o\left(x^{2}\right)\right)\nonumber \\
 & = & x\trace\left(\mathbf{A}\right)-\frac{1}{2}x^{2}\trace\left(\mathbf{A}^{2}\right)+o\left(x^{2}\right)\label{eq:taylor exp}
\end{eqnarray}
The second equation uses Taylor's theorem for several variables at
$\hat{\lambda}_{i}\left(\mathbf{A}\right)=x\lambda_{i}\left(\mathbf{A}\right)$,
since when $x\rightarrow0$, $x\lambda_{i}\left(\mathbf{A}\right)\rightarrow0$
as well. 

Combining (\ref{eq:taylor exp}), (\ref{eq:ebnomin calc}), (\ref{eq:slope calc})
and (\ref{eq:K-user sum rate bound}), we find (\ref{eq:ebnomin})
and (\ref{eq:slope detail-1}).

\section{{\normalsize \label{sec: proof TDMA} }Proof of Theorem {\normalsize \ref{cor:TDMA0}}}

To maximize the right hand side of (\ref{eq:slope detail-1}), we
need to solve the following optimization problem 
\begin{eqnarray}
 & \min_{\hat{\mathbf{V}}_{1},\cdots,\hat{\mathbf{V}}_{K}} & \sum_{j=1}^{K}\left|C_{jj}\right|^{4}\trace\left(\hat{\mathbf{V}}_{j}^{2}\right)\nonumber \\
 &  & +2\sum_{j=1}^{K-1}\sum_{i=j+1}^{K}\left|C_{jj}\right|^{2}\left|C_{ji}\right|^{2}\trace\left(\hat{\mathbf{V}}_{j}\mathbf{U}_{ji}\hat{\mathbf{V}}_{i}\mathbf{U}_{ji}^{\dagger}\right)\label{eq:op prob}\\
 & s.t. & \trace\left(\hat{\mathbf{V}}_{j}\right)=1\nonumber \\
 &  & \hat{\mathbf{V}}_{j}\succeq\mathbf{0}.
\end{eqnarray}

First, consider a simple case where the channel is strictly symmetric:
$\phi_{ji}=0$, $\left|C_{jj}\right|^{2}=1$ and $\left|C_{ji}\right|^{2}=\alpha<1$
for all $i,\, j$. (\ref{eq:op prob}) becomes
\begin{eqnarray}
 & \min_{\hat{\mathbf{V}}_{1},\cdots,\hat{\mathbf{V}}_{K}} & \sum_{j=1}^{K}\trace\left(\hat{\mathbf{V}}_{j}^{2}\right)+2\alpha\sum_{j=1}^{K-1}\sum_{i=j+1}^{K}\trace\left(\hat{\mathbf{V}}_{j}\hat{\mathbf{V}}_{i}\right)\label{eq:op prob-1}\\
 & s.t. & \trace\left(\hat{\mathbf{V}}_{j}\right)=1\nonumber \\
 &  & \hat{\mathbf{V}}_{j}\succeq\mathbf{0}.
\end{eqnarray}
Let the $2\times2$ real positive definite matrix $\hat{\mathbf{V}}_{j}$
be 
\begin{equation}
\hat{\mathbf{V}}_{j}=\left(\begin{array}{cc}
k_{j1} & k_{j3}\\
k_{j3} & k_{j2}
\end{array}\right).\label{eq:k matrix expnd}
\end{equation}
Substituting (\ref{eq:k matrix expnd}) into (\ref{eq:op prob-1}),
we construct a non-linear optimization problem from (\ref{eq:op prob-1})
on standard form: 
\begin{eqnarray}
 & \min_{k_{11},k_{12},k_{13},\cdots,k_{K1},k_{K2},k_{K3}} & \sum_{j=1}^{K}\left(k_{j1}^{2}+k_{j2}^{2}+2k_{j3}^{2}\right)\label{eq:op prob-2}\\
 &  & +2\alpha\sum_{j=1}^{K-1}\sum_{i=j+1}^{K}\left(k_{j1}k_{i1}+k_{j2}k_{i2}+2k_{j3}k_{i3}\right)\nonumber \\
 & s.t. & -k_{j1}\leq0\label{eq:constraint neq1}\\
 &  & -k_{j2}\leq0\label{eq:constraint neq2}\\
 &  & k_{j3}^{2}-k_{j1}k_{j2}\leq0\label{eq:constraint neq3}\\
 &  & k_{j1}+k_{j2}=1\label{eq:constraint eq}\\
 &  & for\, all\, j=1,\cdots,K\nonumber 
\end{eqnarray}
The optimal solution of the problem defined by (\ref{eq:op prob-2})\textasciitilde{}(\ref{eq:constraint eq})
is also the optimal solution of the problem defined by (\ref{eq:op prob-1}).
Denote the optimization problem defined by (\ref{eq:op prob-2})\textasciitilde{}(\ref{eq:constraint eq})
as $\left(P_{\underline{k}}\right)$, where $\underline{k}=\left(k_{11},k_{12},k_{13},\cdots,k_{K1},k_{K2},k_{K3}\right)$
represents the set of feasible solutions. Notice that while any positive
$k_{j1},\, k_{j2}$ with $k_{j1}+k_{j2}\leq1$ satisfies the power
constraint, we require constraint (\ref{eq:constraint eq}) to be
an equality. Because only when it is satisfied with equality, the
system can achieve correct $\ebnominsum$. 

Denote the objective function in (\ref{eq:op prob-2}) by $f\left(\underline{k}\right)$.
Construct the Lagrangian function for problem (\ref{eq:op prob-2})
as 
\begin{eqnarray}
F\left(\underline{k},\,\underline{u}_{1},\underline{u}_{2},\underline{u}_{3},\underline{v}\right) & = & f\left(\underline{k}\right)-\sum_{j=1}^{K}u_{j1}k_{j1}-\sum_{j=1}^{K}u_{j2}k_{j2}\nonumber \\
 &  & +\sum_{j=1}^{K}u_{j3}\left(k_{j3}^{2}-k_{j1}k_{j2}\right)+\sum_{j=1}^{K}v_{j}\left(k_{j1}+k_{j2}-1\right).\label{eq:lagrange}
\end{eqnarray}
To find a optimal solution for this problem, we use Karush-Kuhn-Tucker
(KKT) sufficient condition. It is stated as followed.
\begin{thm}
\label{thm:KKT-Sufficient-Condition}(KKT Sufficient Condition\cite{freund04ocw})
Consider an optimization problem $\left(P\right)$ defined as 
\begin{eqnarray*}
 & \min_{\underline{x}} & f\left(\underline{x}\right)\\
 & \mathrm{subject\; to} & g_{k}\left(\underline{x}\right)\leq0,\, k=1,\cdots,\, m\\
 &  & h_{l}\left(\underline{x}\right)=0,\, l=1,\cdots,\, n,
\end{eqnarray*}
with Lagrangian function 
\begin{eqnarray*}
L\left(\underline{x},\underline{u},\underline{v}\right) & = & f\left(\underline{x}\right)+g\left(\underline{x}\right)^{T}\underline{u}+h\left(\underline{x}\right)^{T}\underline{v}
\end{eqnarray*}
 Let $\underline{x}$ be a feasible solution of $\left(P\right)$,
and suppose $\left(\underline{x},\underline{u},\underline{v}\right)$
satisfy
\begin{eqnarray*}
\nabla_{\underline{x}}L\left(\underline{x},\underline{u},\underline{v}\right) & = & 0\\
\underline{u} & \geq & 0\\
u_{k}g_{k}\left(\underline{x}\right) & = & 0
\end{eqnarray*}
Then if $f\left(\underline{x}\right)$ is a pseudoconvex function,
$g_{k}\left(\underline{x}\right)$, $k=1,\cdots,\, m$ are quasiconvex
functions, and $h_{l}\left(\underline{x}\right)$, $l=1,\cdots,\, n$
are linear functions, then $\underline{x}$ is a global optimal solution.
\end{thm}
Given $\left(P_{\underline{k}}\right)$, it is clear that the objective
function $f\left(\underline{k}\right)$ is a convex function, the
equality constraints (\ref{eq:constraint eq}) are linear, and the
sets of inequality constraints (\ref{eq:constraint neq1}), (\ref{eq:constraint neq2}),
and (\ref{eq:constraint neq3}) are convex. Notice that a convex function
is a special case of pseudoconvex and quasiconvex. Comparing the standard
problem $\left(P\right)$ in Theorem \ref{thm:KKT-Sufficient-Condition}
with our optimization problem $\left(P_{\mathbf{K}}\right)$, we can
conclude that any feasible $\underline{k}$ satisfying 
\begin{eqnarray*}
\nabla_{\underline{k}}F\left(\underline{k},\,\underline{u}_{1},\underline{u}_{2},\,\underline{u}_{3},\underline{v}\right) & = & 0\\
\underline{u}_{1},\underline{u}_{2}\,\mathrm{and}\,\underline{u}_{3} & \geq & 0\\
u_{j1}k_{j1} & = & 0\\
u_{j2}k_{j2} & = & 0\\
u_{j3}\left(k_{j3}^{2}-k_{j1}k_{j2}\right) & = & 0
\end{eqnarray*}
is a global optimal for $\left(P_{\underline{k}}\right)$. Solving
$\nabla_{\underline{k}}F\left(\underline{k},\,\underline{u}_{1},\underline{u}_{2},\,\underline{u}_{3},\underline{v}\right)$
we have
\begin{eqnarray*}
\frac{\nabla F}{\nabla k_{j1}} & = & 2k_{j1}+2\alpha\sum_{i=1,i\neq j}^{K}k_{i1}-u_{j1}-u_{j3}k_{j2}+v_{j}=0\\
\frac{\nabla F}{\nabla k_{j2}} & = & 2k_{j2}+2\alpha\sum_{i=1,i\neq j}^{K}k_{i2}-u_{j2}-u_{j3}k_{j1}+v_{j}=0\\
\frac{\nabla F}{\nabla k_{j3}} & = & 4k_{j3}+4\alpha\sum_{i=1,i\neq j}^{K}k_{i3}+2u_{j3}k_{j3}=0.
\end{eqnarray*}
It is easy to check that $k_{j1}=k_{j2}=\frac{1}{2}$, $k_{j3}=0$
while the Lagrange multipliers $u_{j1}=u_{j2}=u_{j3}=0$, and $v_{j}=-1-\alpha\left(K-1\right)$
satisfy KKT condition. 

Therefore, $k_{j1}=k_{j2}=\frac{1}{2}$, $k_{j3}=0$, i.e. $\hat{\mathbf{V}}_{x_{j}}=\left(\begin{array}{cc}
\frac{1}{2} & 0\\
0 & \frac{1}{2}
\end{array}\right)$ is a global optimal solution. Substitute this optimal solution into
the formula of sum slope (\ref{eq:slope detail-1}), the sum slope
has upper bound 
\begin{eqnarray*}
\slope_{0} & \leq & \frac{2K}{\alpha K+\left(1-\alpha\right)}
\end{eqnarray*}

\section{{\normalsize \label{sec: proof TDMA-1} }Proof of Corollary {\normalsize \ref{cor: TDMA}}}

Before proving this result, we state existing results for general
parametric optimization problems. A general parametric optimization
problem $P\left(\underline{t}\right)$ depending on parameters $\underline{t}\in\mathbb{R}^{r}$
is defined by 
\begin{eqnarray*}
 & \min & f\left(\underline{x},\underline{t}\right)\\
 & \mathrm{subject\, to} & \underline{x}\in\mathbb{R}^{n}\\
 &  & g_{i}\left(\underline{x},\underline{t}\right)\leq0,\, i=1,\cdots,\, s\\
 &  & g_{i}\left(\underline{x},\underline{t}\right)=0,\, i=s+1,\cdots,\, m
\end{eqnarray*}
where $f$ and $g_{i}$ are real functions. Denote the parametric
feasible region by 
\begin{eqnarray*}
A\left(\underline{t}\right) & \triangleq & \left\{ \left.\underline{x}\right|\underline{x}\in\mathbb{R}^{n};\, g_{i}\left(\underline{x},\underline{t}\right)\leq0\, if\, i=1,\cdots,\, s;\right.\\
 &  & \left.g_{i}\left(\underline{x},\underline{t}\right)=0\, if\, i=s+1,\cdots,\, m\right\} .
\end{eqnarray*}
And denote the parametric optimal value function by $\nu\left(\underline{t}\right)\triangleq\inf_{\underline{x}\in A\left(\underline{t}\right)}f\left(\underline{x},\underline{t}\right)$.
The following theorem gives the sufficient condition under which $\nu\left(\underline{t}\right)$
is a continuous function of $\underline{t}$. 
\begin{thm}[Theorem 3, p.70, \cite{aubin2007mathematical}]
\label{thm:(Theorem-3,-p.70,-1} Suppose that 
\begin{enumerate}
\item the function $f$ is continuous on $\underline{x}\times\underline{t}$;
\item the correspondence $A$ is continuous on $\underline{t}$;
\item the subsets $A\left(\underline{t}\right)$ are non empty and compact
\end{enumerate}

Then the optimal value function $\nu\left(\underline{t}\right)$ is
continuous and the correspondence optimal solution set is upper semi-continuous.

\end{thm}
Let $\underline{C}$ correspond to $\underline{t}$, and let the $\underline{k}$
as that defined in Appendix \ref{sec: proof TDMA} correspond to $\underline{x}$
of Theorem \ref{thm:(Theorem-3,-p.70,-1}. It is easy to see that
the objective function of (\ref{eq:op prob}) is continuous on $\underline{k}\times\underline{C}$,
while the feasible region $A\left(\underline{C}\right)$ is non empty,
compact, and independent of $\underline{C}$. Therefore, all three
conditions in Theorem \ref{thm:(Theorem-3,-p.70,-1} are satisfied
and the optimal value function $f\left(\underline{k},\,\underline{C}\right)$
is continuous on $\underline{C}$.

Further, in Theorem \ref{cor:TDMA0} we have shown that when $\underline{C}_{o}=\left\{ \underline{C}:\,\phi_{ji}=0,\,\left|C_{jj}\right|^{2}=1,\,\left|C_{ji}\right|^{2}=\alpha\right\} $
, the optimal value of the objective function of the optimization
problem $P_{\underline{k}}\left(\underline{C}_{o}\right)$ is 
\begin{eqnarray*}
f\left(\underline{k},\,\underline{C}_{o}\right) & = & \frac{2K}{\alpha K+\left(1-\alpha\right)}.
\end{eqnarray*}
Given the continuity of $f\left(\underline{k},\,\underline{C}_{o}\right)$
provided by Theorem \ref{thm:(Theorem-3,-p.70,-1} , for any $\sigma$,
there exist $\sigma_{1},\,\sigma_{2},\,\sigma_{3}$ such that for
the channels $\underline{C}\in\tilde{\mathcal{C}}_{\sigma}$, where
the set $\tilde{\mathcal{C}}_{\sigma}$ is defined as 
\begin{eqnarray*}
\tilde{\mathcal{C}}_{\sigma} & = & \left\{ \underline{C}:\left|\phi_{ji}\right|<\sigma_{1}\right.\\
 &  & \left|\left|C_{jj}\right|^{2}-1\right|<\sigma_{2}\\
 &  & \left|\sqrt{\left|C_{ji}\right|^{2}}-\alpha\right|<\sigma_{3}\\
 &  & \nicefrac{\left|C_{ij}\right|^{2}}{\left|C_{jj}\right|^{2}}<1\\
 &  & \left.\underline{C}\in\mathcal{C}_{\alpha}\right\} ,
\end{eqnarray*}
the optimal value of the objective function of the optimization problem
$P_{\underline{k}}\left(\underline{C}\right)$ satisfies 
\begin{eqnarray*}
\left|f\left(\underline{k},\,\underline{C}\right)-f\left(\underline{k},\,\underline{C}_{o}\right)\right| & < & \sigma.
\end{eqnarray*}
Notice that $\mathcal{C}_{\alpha}$ is defined in Theorem \ref{thm:K-user sum rate}. 

Because $\mathbf{1}\in\mbox{cl}\left(\tilde{\mathcal{C}}_{\sigma}\right)$,
as $\alpha\rightarrow1$, for any positive $\sigma$, there exists
$\tilde{\mathcal{C}}_{\sigma}$, such that for $\underline{C}\in\tilde{\mathcal{C}}_{\sigma}$
its sum slope satisfies 
\begin{eqnarray}
\slope_{0} & \leq & 2+\sigma,\label{eq:almost TDMA-1}
\end{eqnarray}
If the magnitude and phase of the channel coefficients are drawn from
continuous random distribution, $Pr\left(\tilde{\mathcal{C}}_{\sigma}\right)>0$. 

And as $\sigma\rightarrow0$, 
\begin{eqnarray*}
\lim_{\sigma\rightarrow0}\Delta\slope_{0} & = & \frac{1}{K}
\end{eqnarray*}

\end{document}